\documentclass[a4paper,twoside]{article}
\newcommand{\affil}[1]{$^{\rm #1}$}
\usepackage[authoryear]{natbib}
\usepackage{multirow}
\bibpunct{(}{)}{;}{a}{}{,}
\usepackage{graphicx}
\usepackage{url}
\date{} 
\title{\large\bf\flushleft Are tiled display walls needed for astronomy?}
\author{\parbox{\textwidth}{\flushleft
\vspace{-0.5cm}
{\it Bernard F.\ Meade\affil{A,B,C}, Christopher J.\ Fluke\affil{A}, Steven Manos\affil{B} and Richard O.\ Sinnott\affil{B}}\\
\vspace{0.4cm}
{\small \affil{A}\,Centre for Astrophysics and Supercomputing, Swinburne University of Technology, PO Box 218, Hawthorn, Australia, 3122}\\
{\small \affil{B}\,The University of Melbourne, Parkville, Victoria, Australia, 3010}\\
{\small \affil{C}\,Email: bmeade@unimelb.edu.au}}}
\begin{document}
\twocolumn[
\begin{changemargin}{.8cm}{.5cm}
\begin{minipage}{.9\textwidth}
\vspace{-1cm}
\maketitle
%
%
\small{\bf Abstract:} 
Clustering commodity displays into a Tiled Display Wall (TDW) provides a cost-effective way to create an extremely high resolution display, capable of approaching the image sizes now generated by modern astronomical instruments.  Many research institutions have constructed TDWs on the basis that they will improve the scientific outcomes of astronomical imagery.  We test this concept by presenting sample images to astronomers and non-astronomers using a standard desktop display (SDD) and a TDW.  These samples include standard English words, wide field galaxy surveys and nebulae mosaics from the Hubble telescope. Our experiments show that TDWs provide a better environment than SDDs for searching for small targets in large images. They also show that astronomers tend to be better at searching images for targets than non-astronomers, both groups are generally better when employing physical navigation as opposed to virtual navigation,  and that the combination of two non-astronomers using a TDW rivals the experience of a single astronomer. However, there is also a large distribution in aptitude amongst the participants and the nature of the content also plays a significant role in success.


\medskip
\medskip
\end{minipage}
\end{changemargin}
]
\small
\section{Introduction}
\label{sct:introduction}
Astronomy produces some of the largest volumes of scientific data.  Future facilities such as the Large Synoptic Survey Telescope \citep{tyson2002, ivezic2008} and the Square Kilometer Array (SKA)\footnote{\protect\url{http://www.skatelescope.org}} will produce final datasets heading toward, or even beyond, exabyte sizes.

In this `big data' era of astronomy, existing data analysis tools and methodologies, where the astronomer works directly on the data at the desktop, will be pushed to their limits. There will be an ever-increasing reliance on automated processes to identify objects of interest.  This includes the growing variety of generic approaches referred to as data mining \citep{ball2010, brescia2012, way2012}, and discipline specific solutions such as automated source finders [e.g. see \citep{koribalski2012} for a recent review of HI source finding strategies].   

As valuable as automatic analyses of these enormous datasets are, astronomy still relies heavily on visual inspection.  As the sensitivity of telescopes and detectors is improved, phenomena are increasingly being revealed at the boundary between the signal and the noise.  In many cases, these phenomena are not even predicted, making automatic analysis meaningless.  It is often a case of not knowing what you are looking for until you see it \citep{hassan2011}.

Not only is the total volume of astronomy data increasing, but the size of individual images (and data cubes) is growing as well.    For example, one of the highest resolution cameras currently available is the Dark Energy Camera (DECam), part of the Dark Energy Survey \citep{darkenergysurveyweb}. This camera uses an array of 62 x 2048x4096 CCDs to form a 520 Megapixel image \citep{mohr2012}.   However, as Table \ref{tbl:rescomp} demonstrates, there is a growing divide between the resolution of images that can be recorded and the resolution of images that can be displayed on the desktop. 
\begin{table*}
\caption{Comparison between the typical displays available to an astronomer, and the resolution of some of the current and proposed astronomical cameras.  MPs = Megapixels.}
\label{tbl:rescomp}
\begin{center}
\begin{tabular}{llll}
{\bf Capture Device}  & {\bf Resolution} & {\bf MPs} & {\bf Reference} \\
\hline
HST Advanced Camera for Surveys &2x2048x4096	& 16 & \citep{acsweb} \\
Skymapper		&32x2048x4096 & 268 & \citep{keller2007} \\ 
DECam	 &62x2048x4096	& 520 & \citep{mohr2012} \\ 
Subaru Hyper Suprime-Cam	 &104x2048x4096	 & 870 & \citep{hypersuprimecamweb}\\
\hline
{\bf Display Device}	 &  {\bf Resolution} &	{\bf MPs } \\
\hline
Standard desktop display 	&1680x1050	&1.7\\
Full high-definition (FHD) desktop display	 &1920x1200	&2.3\\
iPad (with retina display)	 &2048x1536	&3.1\\
Dell UltraSharp desktop display 	& 2560x1600	&4.1\\
Laptop (Macbook Pro)	 & 2880x1800	&5.2\\
4K ultra-high definition (UHD) display	 & 3840x2160&	8.3\\
\hline
\end{tabular}
\end{center}
\end{table*}

When exploring astronomical imagery, it is desirable to display an image at its native resolution, where there is a one-to-one correspondence between image and display pixels.  A very large display with low resolution may reveal less information than a smaller display with a higher resolution. Such high-resolution images reveal more than just the detail of individual celestial objects.  In fact, it is the combination of detail and context that make these images valuable: understanding the environment is critical to describing the phenomenon itself.    When the image dimensions exceed the capabilities of a standard desktop display, then it is time to look to a non-standard display such as a tiled display wall. 

\begin{figure*}[t]
\begin{center} 
\includegraphics[width=16cm, angle=0]{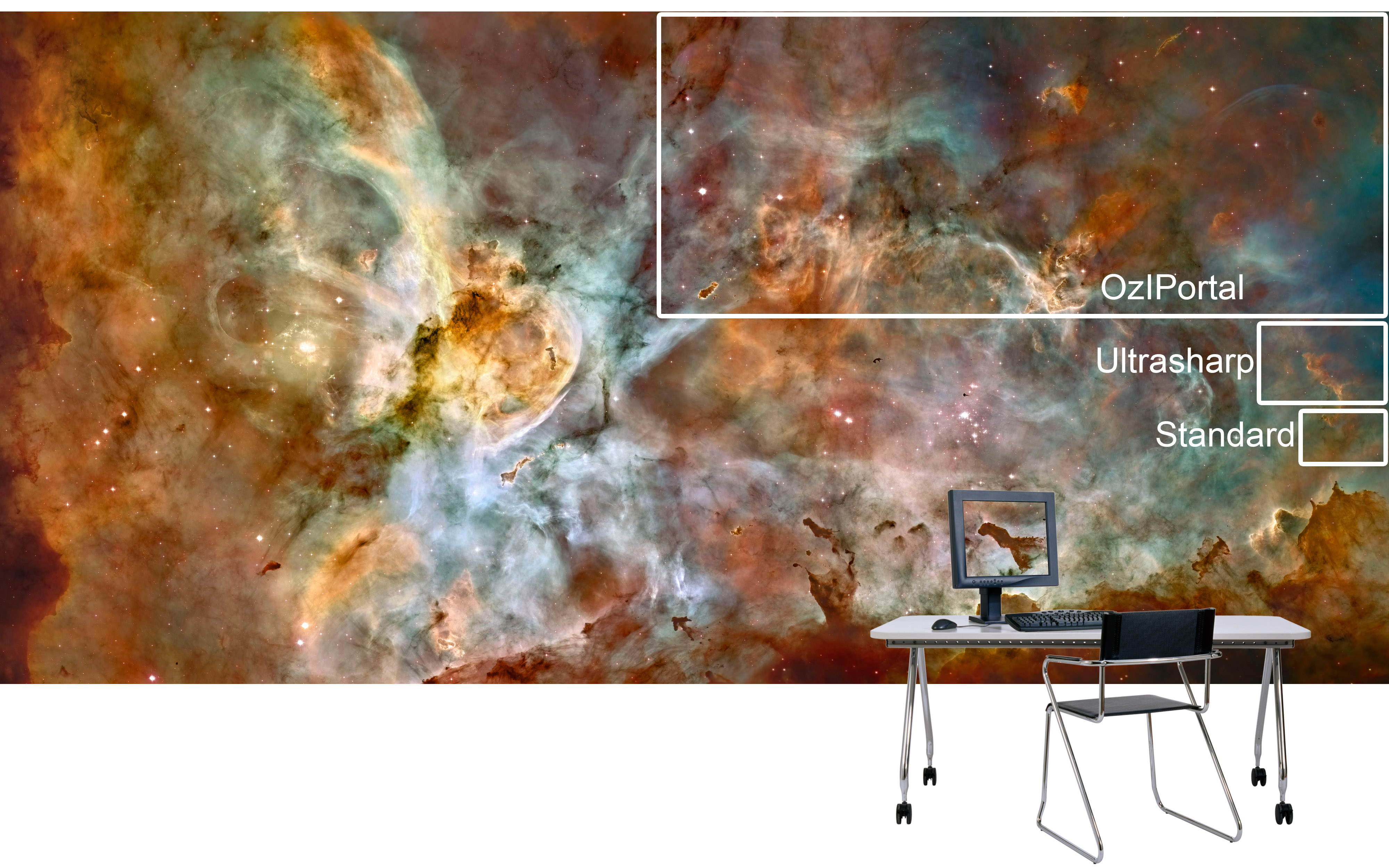}
\caption{The 29566 x 14321 pixel Carina Nebula mosaic from Hubblesite.org, with OzIPortal (15360x6400), Dell Ultrasharp (2560x1600) and Standard Desktop Display (1680x1050) sizes overlaid. }
\label{fig:carinascales}
\end{center}
\end{figure*}

\subsection{Tiled Display Walls}
\label{sct:tdw}
A tiled display wall is an ultra-high-resolution display comprising a two-dimensional matrix of lower resolution display components, typically standard flat-screen monitors.   While there are some slight differences in the way specific tiled display walls are assembled and configured, e.g. Hiperwalls \citep{hiperwallweb}, Powerwalls \citep{leedspowerwallweb}, or OptIPortals (see Appendix \ref{sct:optiportalproject}), they still operate in a similar manner and hereafter are described simply as TDWs.

A key element of the design principle of TDWs is the use of commodity computers and displays.  The computing power available in a standard desktop device with a typical graphics card is capable of driving stunning graphics across multiple displays at very high frame rates.  Similarly, the expansion of capabilities of devices such as the emergence of multi-head graphics cards and additional expansion slots on motherboards, means that a single computer can now drive many displays.  In fact, a modern computer containing a motherboard with three PCI-Express slots, each hosting a dual-head graphics card, with six Matrox TripleHead2Go\footnote{\protect\url{http://www.matrox.com/graphics/en/products/gxm/}}devices on each output, can drive 18 full high definition (FHD) displays.

While most TDWs are designed and built as flat screens, either free-standing or mounted on a wall, the use of individual display elements provides a great deal of flexibility in the geometrical configuration. The Mechdyne CAVE2 systems at the Electronics Visualisation Lab (EVL) (University of Chicago) \citep{febretti2013} and Monash University, Australia, wrap the TDW around the user, providing an extremely high-resolution immersive stereoscopic environment (74 Megaixels in 2D or 37 Megapixels in 3D).  Two key advantages of using monitors over large-screen rear-projection, the usually approach for Cave Automatic Virtual Environments \citep[CAVE;][]{cruz1992}, is the increase in both the available pixels and the display brightness.  A third advantage is the great reduction in physical footprint of the facility compared to the CAVE, which requires extra space outside of the walls to house the data projectors.  The trade-off is a more complex computing and network back-end to drive $\sim 80$ individual panels, rather than the (maximum) six walls of  a cubic CAVE.  Additionally,  there is the visible presence of screen bezels - the frame around each of the display elements.

While an ideal TDW would provide a seamless image, in reality the screen bezels introduce a windowing effect.  Bezels can be distracting for certain types of content (e.g. office applications), whereas for other tasks they can actually provide a natural coordinate grid to aid in exploration (see Section \ref{sct:feedback}).   The display panels themselves continue to improve, including the appearance of screens with very thin bezels, such as the Christie Digital FHD551-X \footnote{\protect\url{http://www.christiedigital.com/en-us/digital-signage/products/lcd-flat-panels/pages/55-hd-lcd-flat-panel.aspx}}with only 5.5mm combined bezel width.

We distinguish between resolution and pixel density when displaying images.  For example, the first release of the Retina display for the 15 inch Apple Macbook Pro\footnote{\protect\url{http://www.apple.com/au/macbook-pro/features-retina/, 2013}} had a resolution of 2880x1800, which greatly exceeds the typical resolution of a FHD home theatre display at 1920x1080.  However, the home theatre's 2 megapixel display can extend over 150 inches (measured diagonally), while the Macbook Pro display crams its 5.1 megapixels into a 15 inch screen.  The pixel densities of each configuration are at the extremes, with the Macbook Pro Retina providing a practically seamless image, while the FHD image projected to 200 inches would reveal the individual pixels quite clearly.  At this time, there remains a significant price jump to move from FHD to the next off-the-shelf resolution of 2560x1600 pixels (e.g. Dell Ultrasharp).  However, the recent emergence of commercially available 4K systems and the Retina displays from companies like Panasonic and Apple, will likely drive down the price of the 2560x1600 displays.

A number of applications exist to simplify the management of a TDW.  The two main contenders are Scalable Adaptive Graphics Environment \citep[SAGE;][]{sageweb} from the University of Chicago's EVL and the Cross-platform Cluster Graphics Library \citep[CGLX;][]{cglxweb, ponto2011} from UC San Diego's CALIT2, though several other solutions also exist.  One of the principle benefits of SAGE is that it makes sharing content between TDWs easy \citep{fuliwara2011}.  More recently, \citet{tada2011} have developed a visualization adaptor that extends the capability of SAGE to allow the display of any X-Window, which opens up the possibility of using almost any application on the TDW.

\subsection{Background}
\label{sct:background}
Astronomical imagery is often seen on promotional material for TDWs, such as the SAGE gallery images\footnote{\protect\url{http://http://www.sagecommons.org/community/sage-walls/, 2013}}.  Indeed, the very high resolution images captured by modern detectors (Table \ref{tbl:rescomp})  do seem very well suited to the environment.   TDWs have been used successfully as public outreach devices, such as the displays at the National Institute of Information and Communications Technology, Japan \citep{morikawa2010, nictweb}, and the Adler Planetarium, USA \citep{adlerweb}.  However, there is a paucity of literature examining whether TDWs actually do improve understanding of any ultra-high resolution image in any scientific discipline.   
 
\citet{ball2005a} conducted some of the first experiments to compare individual computer displays with TDWs. They tested subjects on a single desktop display (17 inch, 1280x1024 pixels), a 2x2-display TDW (2560x2048 pixels) and a 3x3-display TDW (3840x3072 pixels). For a target search (red shapes on a black background with random grey dots), they found that participants performed far better when searching for small targets when they could see all the targets at once.  The first part of the experiment involved the participants searching for a specific configuration of red dots, while the second part required the participants to match pairs of configurations.  No statistically significant difference was observed for targets that could be easily seen on all display environments without needing to pan or zoom.
 
\citet{ball2005a}  also found that the experience of virtual navigation, that is, using a mouse to zoom and pan on a single display, caused more frustration for the participants than the physical navigation required for the TDW.  Here, physical navigation means observing parts of an image by physically moving the eyes, head or whole body to an optimum position.  The more virtual navigation that was required, the greater the disorientation and agitation experienced by the participants.  The authors suggest that being able to easily maintain context while searching for detail made the TDW a more acceptable experience.  They also found that the physical construction of the TDW with the screen bezels dividing the image into segments aided the search process.  We comment on this issue in Section \ref{sct:feedback}.
 
\citet{yost2007} studied the visual acuity of human perception with regards to high-resolution displays.  In this context, visual acuity refers to the ability to perceive all displayed information:  when an entire display surface is within the user's field of view, and the individual pixels remain discernible, the display is said to be within visual acuity.  While increasing the pixel density is one way to exceed visual acuity, increasing both the pixel count and the display size is another.  When a display exceeds visual acuity, there are always pixels that cannot be accurately perceived without physical navigation.  However, increasing physical navigation does not negatively impact on the performance on most tasks, whereas virtual navigation of the same data or image does have a significant negative impact.  This is contrary to the notion that there is no value in using a display that exceeds visual acuity.  For practical reasons, it is harder to establish at what point this advantage disappears.
 
\citet{andrews2011} focused their investigation on the human experience of using high-resolution displays.  In this study, they defined large displays as being human-scale, that is, where the physical size of the display was of similar height and width to naturally occupy the natural field of view of an adult human.  In this definition of a large, high-resolution display, this can be achieved through tiling of displays or using individual displays with greater pixel count and large physical size.  The authors argue that the design of such displays as TDWs would greatly benefit from considering the physical nature of human-centric search techniques and creating displays that meet these needs.
 
\citet{andrews2011} also considered the natural perceptions of users and how a display can affect these perceptions.  There is a potential for TDWs to overwhelm the user with information, or induce physical fatigue due to the increased requirements of physical navigation of the environment.  However other studies have shown that physical navigation can outperform virtual navigation for many tasks \citep{ball2007}, and that users quickly adjust to the information density shown on a large display \citep{andrews2010}.  The increased physical activity required for a large display has not shown significant increase in fatigue of subjects, though there is a possibility of some discomfort in the neck due to the increased turning of the head \citep{ball2005b, bi2009}. 
 
Following on from these design questions, \citet{bezerianos2012} conducted experiments to determine how proximity to a TDW could affect the perceptions of the user, particularly focusing on angular distortion effects.  The ability of participants to effectively estimate quantities such as angle, area and length of objects within an image were significantly affected when the angle of presentation was increased.  Thus when a participant was very close to the TDW, their ability to estimate these quantities diminished as the distance of the object to the subject increased.  Of particular interest are the results from the second experiment where some of the participants were required to remain in a fixed location, while others were allowed to move freely.  The study found that the static position yielded just as accurate results of the mobile position, but was less time consuming.  Therefore, the authors' recommendation that users be encouraged to remain at a distance from the TDW where possible, or physically inspect objects positioned close to their position, ties in closely with the {ball2005a}  observation that users naturally avoid using virtual navigation unless they absolutely have to \citep{ball2007}.
 
Most of these studies focus on generalized examples of use of TDW, but the nature of the research disciplines also needs to be considered when investigating these displays.  For example, the way an astronomer would use a TDW could have significant differences to the way an economist would use it.  As \citet{moreland2012} argues, we already know how to build the displays, but we have little experience in considering domain-specific applications.  The desirability of achieving a one-to-one correspondence between image and display pixels aside, it is far too simplistic to suggest that images of A x B pixels require displays of equivalent resolution.  Instead, the need must be borne out of the research and the data, where the impact of virtual navigation impedes comprehension.

\subsection{Overview}
\label{sct:overview}
In this paper, we describe a series of experiments designed to investigate the assumption that TDWs are intrinsically beneficial in astronomical research.  We focus our attention on targeted searches within high resolution images that exceed the available resolution of  a standard desktop display.  We consider the performance of both individuals and pairs of users at finding targets of decreasing size on either a standard desktop display or a TDW.  The participants in the experiments included professional astronomers, experienced amateur astronomers and non-astronomers.

The TDW used in these experiments, the OzIPortal, was built by the School of Engineering at the University of Melbourne in 2008 and is now operated by the University's central IT department.   The TDW comprises a 6x4 matrix of Dell Ultrasharp monitors (2560x1600).  With a total resolution of 15360x6400 pixels, it is capable of displaying 98.3 Megapixels.  However, as Figure~\ref{fig:carinascales} shows, this is less than a third of the pixels available in images such as Hubble's Carina Nebula mosaic\footnote{\protect\url{http://imgsrc.hubblesite.org/hu/db/images/hs-2007-16-a-full\_jpg.jpg}}. 

The OzIPortal initially used CGLX for the interface, but this was replaced with the somewhat more versatile SAGE software. We describe the history of the OzIPortal in Appendix \ref{sct:oziportal}.

The remainder of this paper is set out as follows. In Section \ref{sct:experiments}, we describe the OzIPortal experiments, including the image selections, participants and procedure.  In Section \ref{sct:results}, we show the experimental results.  We look at the comparative performance of targeted searches using both standard desktop display and TDW environments. We consider the performance of the non-astronomer, astronomer and collaborative pair groups. We comment on key findings from the post experiment survey and video observations.  In Section \ref{sct:discussion} we discuss the implications of these results in the context of the potential use of TDWs in  astronomy.  We consider further experiments that are either extensions of the current work, or alternative aspects of using a TDW that might be beneficial to astronomers.  Concluding remarks are made in Section \ref{sct:conclusion}.

\section{The OzIPortal experiments}
\label{sct:experiments}
In this section, we describe our experimental procedure to investigate the role TDWs might play in aiding  knowledge discovery and comprehension of ultra-high resolution images (i.e. $\sim 100$ megapixels).  These images provide researchers with an opportunity to seamlessly explore both context and detail at will.  Yet on a standard desktop display (SDD), defined for our purposes as a 24 inch LCD with a resolution of 1680 x 1050 pixels\footnote{The recommended size for centrally deployed computers at the University of Melbourne at the time of the experimental work}, a researcher must choose dynamically between context or detail, as both cannot be seen at once.   In particular, we wanted to determine if there was indeed a definable performance improvement when using a TDW compared to a SDD, which corresponds to the popular expectation that big images need a big display to be seen ``properly''.  The high-resolution images and target objects were chosen from three different categories: English words, galaxies and nebulae (see section \ref{sct:imageselection}).

\begin{figure}
\begin{center} 
\includegraphics[width=7.5cm, angle=0]{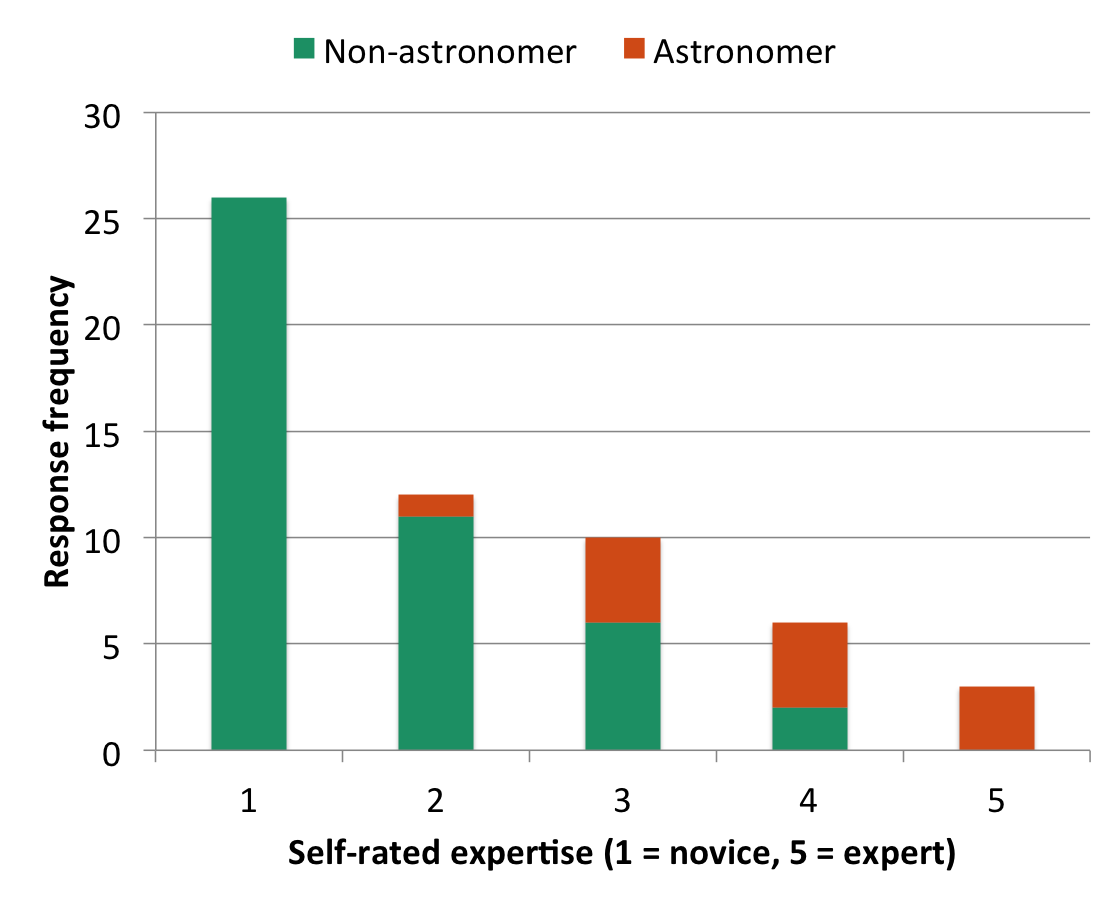}
\caption{Survey results for self-rated level of expertise with astronomical imagery, for the astronomer and non-astronomer groups.  The non-astronomer cohort (green) self-identified strongly with the low end of the experience scale, while the astronomer group (red) is towards the high end.  }
\label{fig:selfrate}
\end{center}
\end{figure}

\subsection{Participant selection}
\label{sct:participants}
Participants were recruited from two different demographic categories:  astronomers and non-astronomers.  For the astronomers, participants included academics, postdoctoral researchers, research students and advanced amateurs.  Within this group there was a mix of radio, optical and theoretical astronomers.    Non-astronomer participants had a wide range of experience with astronomical imagery, ranging from none to a high level of familiarity.   As such a secondary category of expert and non-expert was introduced, based on the participant's self-rated level of experience with astronomical imagery.  Figure \ref{fig:selfrate} shows that the non-astronomer cohort self-identified strongly with the low end of the experience scale, while the astronomer group is towards the high end.   This self-rating reflects that, for example, a theoretical astronomer may not feel they have the same expertise as an optical astronomy who works constantly with images.  

A total of 45 non-astronomers and 12 astronomers participated in a range of experiments. All participants had a reasonably high-level of familiarity with graphical user interfaces and the use of a mouse for panning and zooming, but few had any prior exposure to a TDW.  We report here on the performance results of a subset of 30 participants, noting that:
\begin{itemize}
\item The first five non-astronomer subjects participated in an experiment refinement phase and thus their performance results have been excluded from the results described below.
\item 14 non-astronomers were presented with a  slightly revised set of tasks to those described here.  These additional tasks focused on a small target search and multiple image inspection. The small target search proved too difficult to complete in the SDD environment due to a "too-restrictive" time limit of two minutes, and too few participants were available to complete the multiple image inspection. 
\end{itemize}
As all of these participants did complete the post-experiment survey, providing relevant comments on issues such as the suitability of the TDW for the target search task, we retained their survey responses for subsequent qualitative interpretation.

16 of the remaining non-astronomers completed a target search in pairs in order to investigate the process of collaborative inspection on SDDs and TDWs (see Section \ref{sct:collaboration}).

\begin{figure*}
\begin{center} 
\includegraphics[width=16cm, angle=0]{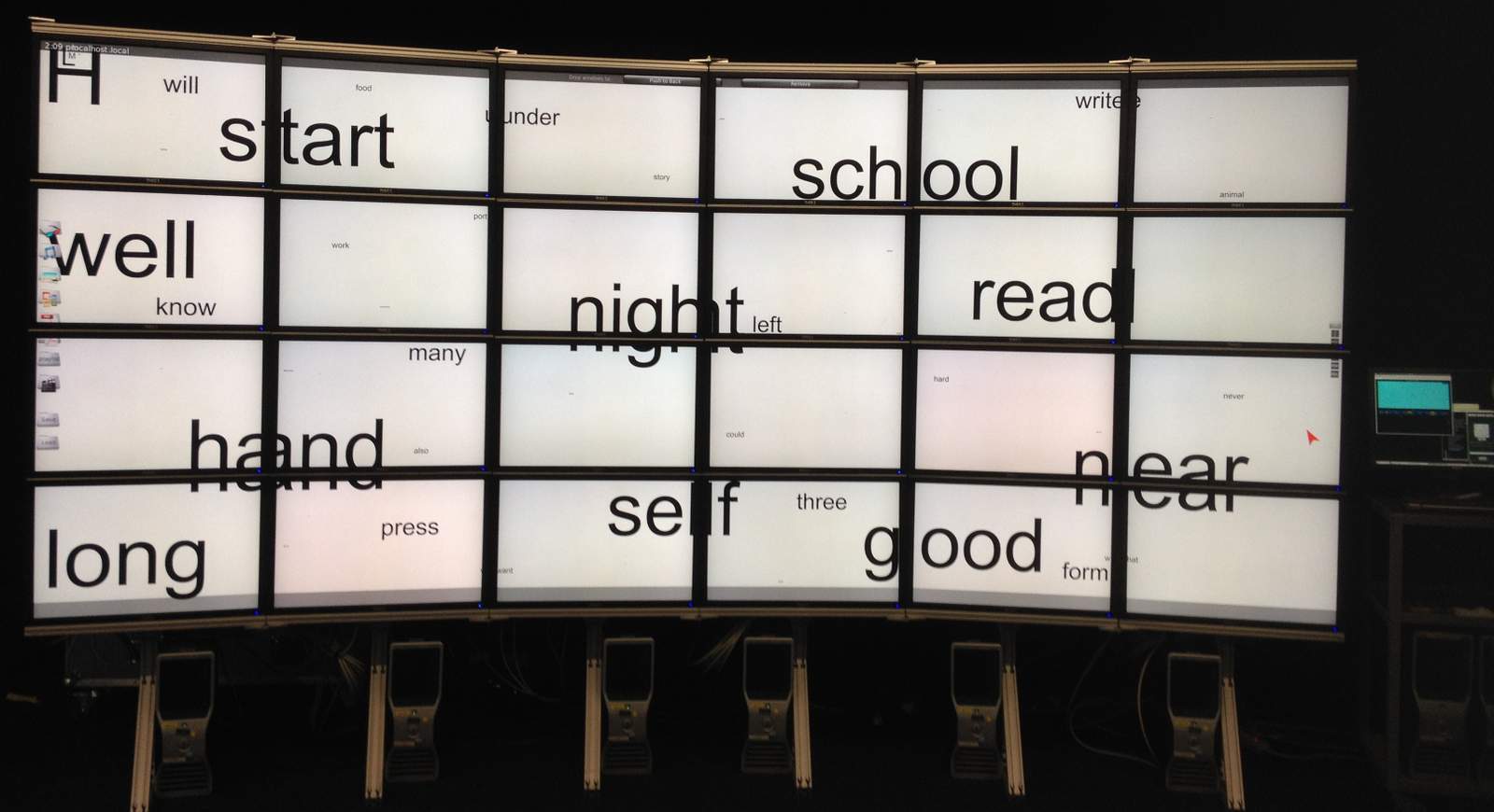}
\caption{The OzIPortal TDW with English word targets displayed at their native resolution. Arial font sizes used were 1000, 300, 100, 30, and 10 points.  All words were visible on the TDW using physical navigation and no zooming. }
\label{fig:wordstdw}
\end{center}
\end{figure*}

\begin{figure*}[t]
\begin{center} 
\includegraphics[width=16cm, angle=0]{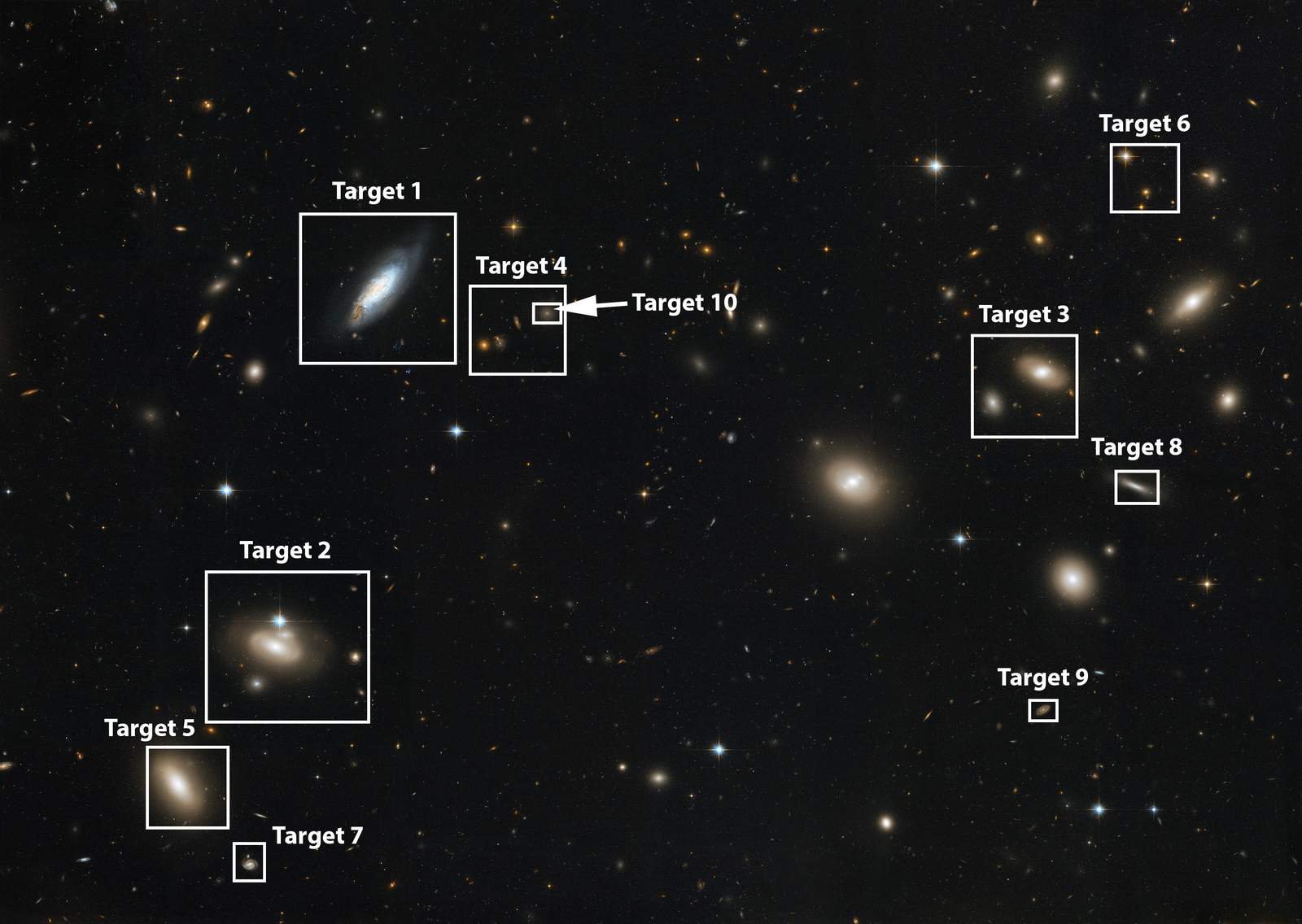}
\caption{Galaxy Set A targets in the Coma Cluster ({\tt\ \protect\url{http://hubblesite.org/newscenter/archive/releases/2008/24/image/a/}})}
\label{fig:galtargetsA}
\end{center}
\end{figure*}

\begin{figure*}
\begin{center} 
\includegraphics[width=16cm, angle=0]{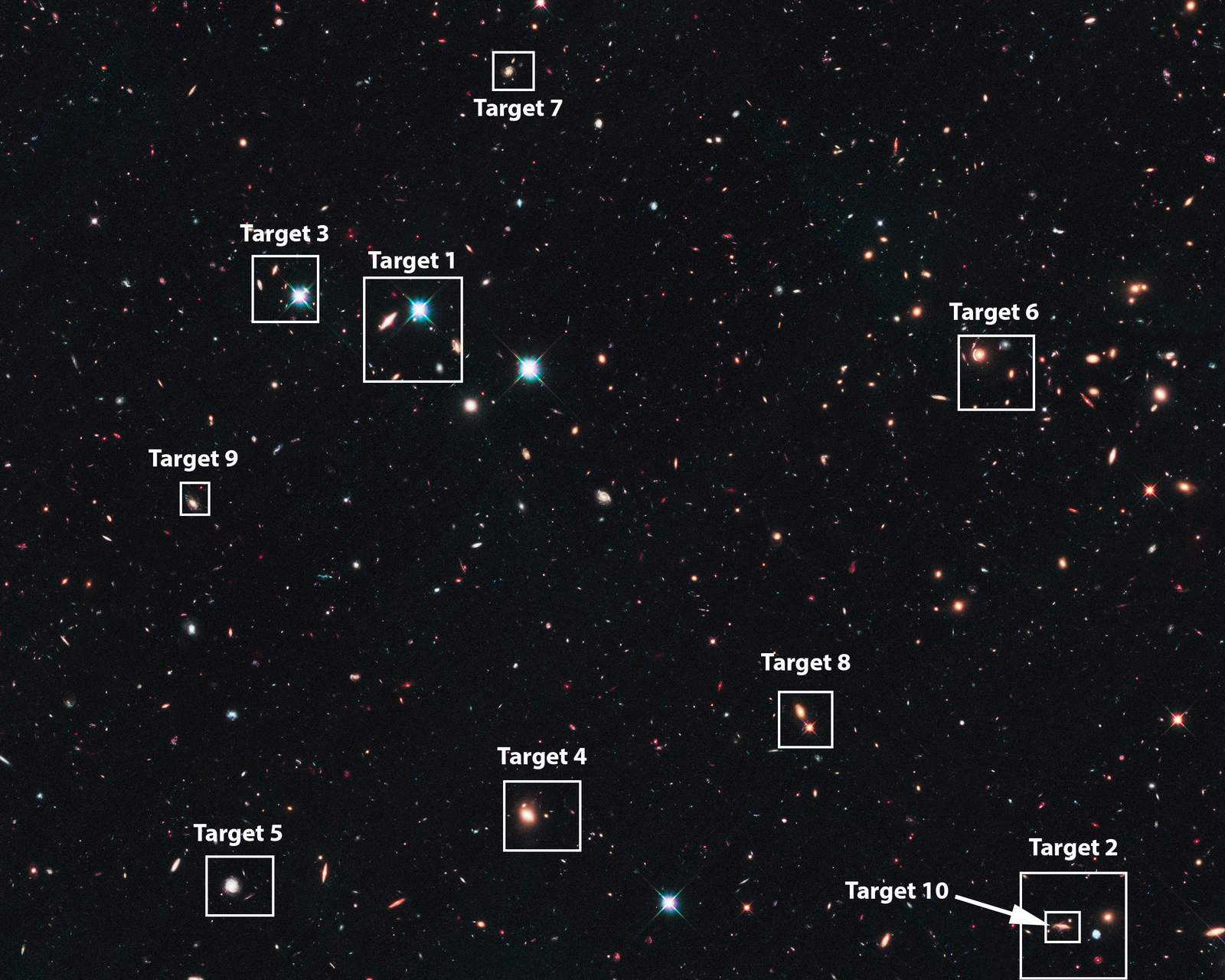}
\caption{Galaxy Set B targets in the CANDELS Ultra Deep Survey ({\tt \protect\url{http://hubblesite.org/gallery/album/entire/pr2013011b/hires/true/}})}
\label{fig:galtargetsB}
\end{center}
\end{figure*}

\begin{figure*}
\begin{center} 
\includegraphics[width=16cm, angle=0]{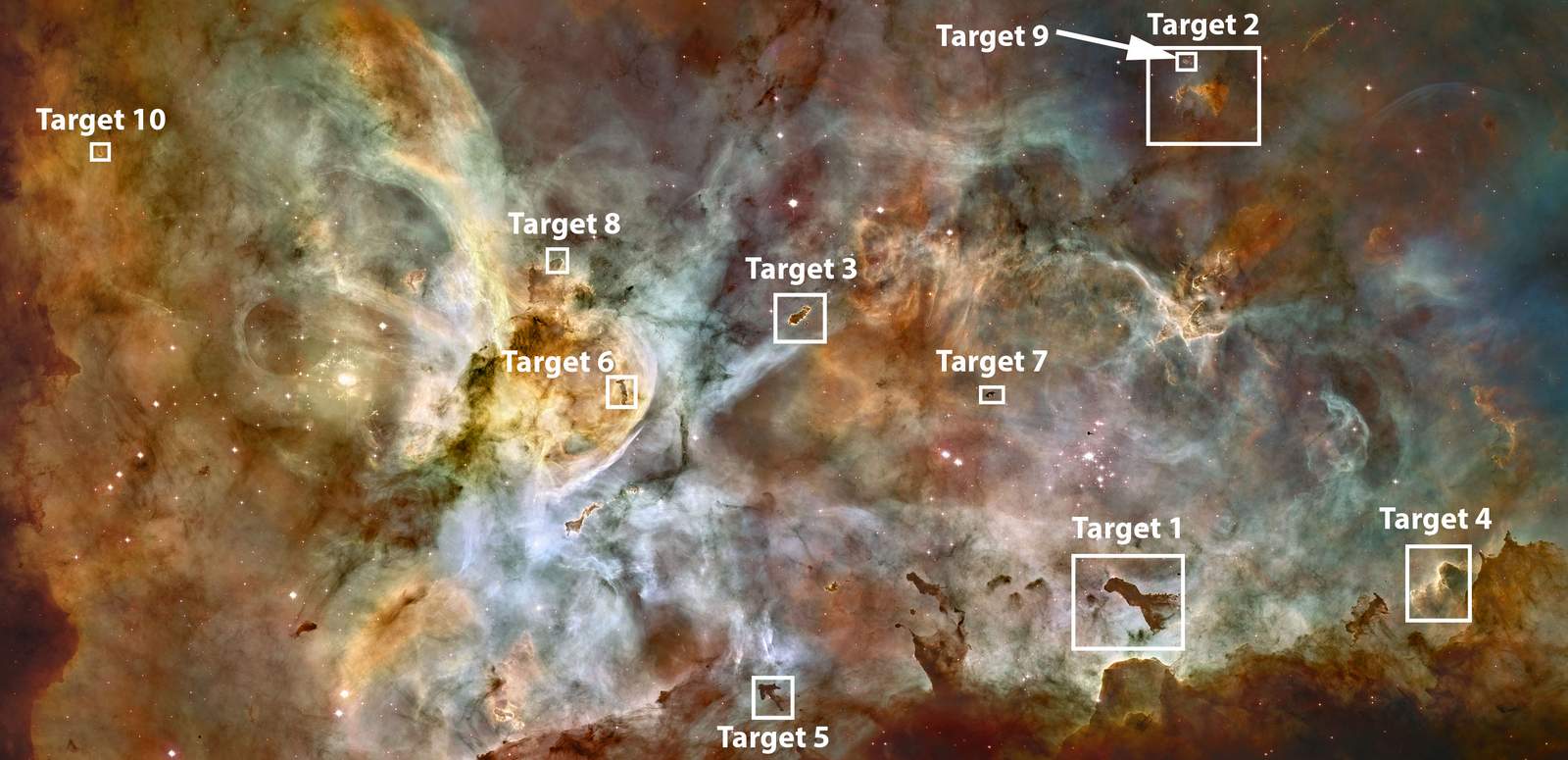}
\caption{Nebula Set A targets in the Carina Nebula ({\tt \protect\url{http://hubblesite.org/gallery/album/nebula/pr2007016a/hires/true/}})}
\label{fig:nebtargetsA}
\end{center}
\end{figure*}

\begin{figure*}
\begin{center} 
\includegraphics[width=16cm, angle=0]{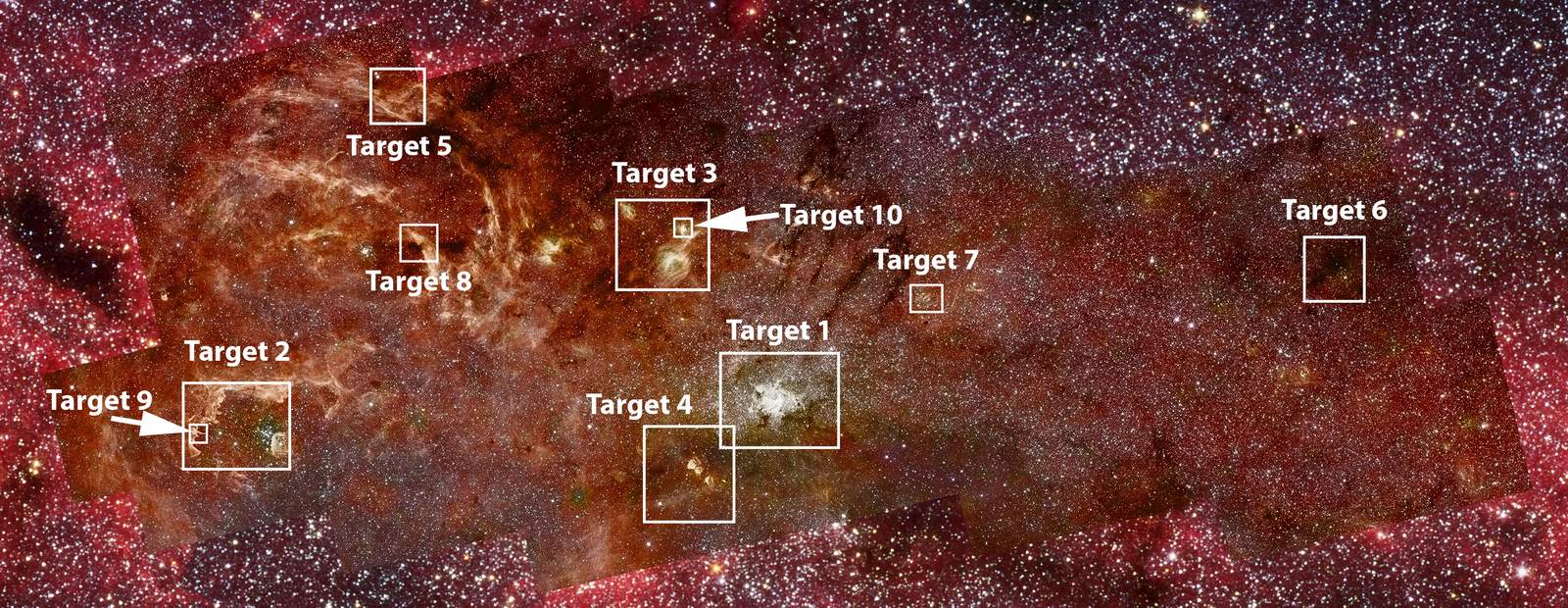}
\caption{Nebula Set B targets in the HST-Spitzer Galactic Center composite ({\tt \protect\url{http://hubblesite.org/newscenter/archive/releases/2009/02/image/d/}})}
\label{fig:nebtargetsB}
\end{center}
\end{figure*}

\begin{table*}[t]
\caption{Images used for the galaxy search and nebula search.  Note that the Carina Nebula image was displayed at 50\% of the native resolution for performance reasons}
\label{tbl:images}
\begin{center}
\begin{tabular}{lllll}
{\bf Image ID} & {\bf Field description} & {\bf Resolution} & {\bf Targets} \\ \hline
Galaxy Set A & The Coma Cluster & $10816 \times 7679$ & Figure \ref{fig:galtargetsA} \\
Galaxy Set B & CANDELS Ultra Deep Survey & 15516 x 8255  & Figure \ref{fig:galtargetsB} \\
Nebula Set A &  The Carina Nebula (NGC3772) & 29566 x 14321 @ 50\% & Figure \ref{fig:nebtargetsA} \\
Nebula Set B & HST-Spitzer Composite of Galactic Center & 12203 x 4731 & Figure \ref{fig:nebtargetsB}
\end{tabular}
\end{center}
\end{table*}

\subsection{Image and target selection}
\label{sct:imageselection}
In order to establish a common ground between the astronomer and non-astronomer groups, the first image  was made up of black words on a white background at a resolution that precisely matched the TDW (15360 x 6400), as can be seen in Figure \ref{fig:wordstdw}.  At this resolution, all words were readable on the TDW without the need to zoom the image.  

The words were taken from a list of the top 250 English words \citep{anglikweb} to ensure all participants were familiar with the targets.  The words were rendered in Arial font and were sized in points of 1000, 300, 100, 30, and 10.  Five targets were created at each size except the smallest where an extra five were added.  For each size, an equivalent number of non-target words were added from the same list, to reduce the possibility of participants guessing based purely on size.

When viewed on the SDD, scaling the image to full screen reduced readability to words in 100pt font or greater.  For words 30pt or 10pt in size, zooming the image was necessary.

The astronomy targets were chosen to present a range of sizes similar to the word sizes described above, chosen from amongst the largest available on the HubbleSite gallery\footnote{\protect\url{http://hubblesite.org/gallery/album/entire/hires/true/}} - see Table \ref{tbl:images} for details.  For performance reasons, the Carina Nebula mosaic was shown at 50\% of the native resolution.

From these images, the search targets were selected to roughly correspond to the physical sizes of the words, without following a strict sizing scale.  The largest astronomy target was 2100 x 1730 pixels while the smallest target was 185 x 145 pixels.  Images were not rotated, but were scaled to appear the same size on the search target presentation screen. Astronomical targets were chosen to reflect increasing difficulty. Due to the increased difficulty of the astronomical search compared to the word search, and the limited amount of time available for each participant to complete each task, the number of targets was restricted to 10 per image.

Targets selected from the galaxy images included structures around the galaxy.  However, these targets exist on a black background and have no visible connectivity to the other objects in the image.  Nebulae provide a fully connected structure with details visibly connected to the context.  Figures 
\ref{fig:galtargetsA} to \ref{fig:nebtargetsB} show each of the astronomical images and the targets.  Additional galaxy and nebula images were used to introduce the environments but were not used during the experiment.

In order to eliminate any potential presentation bias, the image sets were shown alternating for the environments, so that half the participants saw the set A images on the SDD and set B images on the TDW, and vice-versa for the rest.

\begin{figure*}
\begin{center} 
\includegraphics[width=16cm, angle=0]{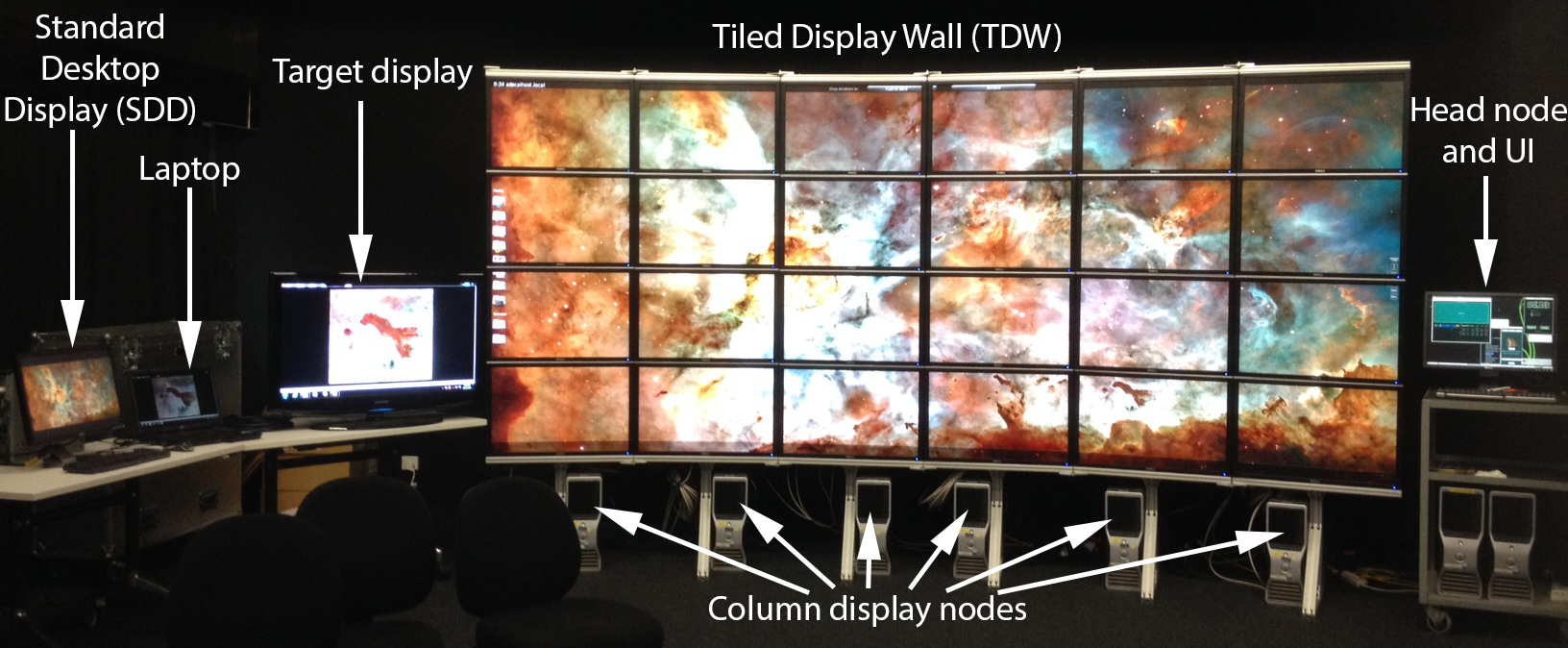}
\caption{Experiment layout as described in Section \ref{sct:procedure}. The individual target objects were presented on a 40 inch LCD TV (Target Display) immediately adjacent to the TDW, as well as on a laptop sitting adjacent to the SDD (Laptop).  A standard Microsoft PowerPoint presentation was used to display the targets to the participant.  The six columns of the OzIPortal are driven by six column display nodes, with master control under the SAGE environment from the Head node [Image: Carina nebula mosaic from \protect\url{http://www.hubblesite.org].}}
\label{fig:exptsetup}
\end{center}
\end{figure*}

\begin{figure}
\begin{center} 
\includegraphics[width=7.5cm, angle=0]{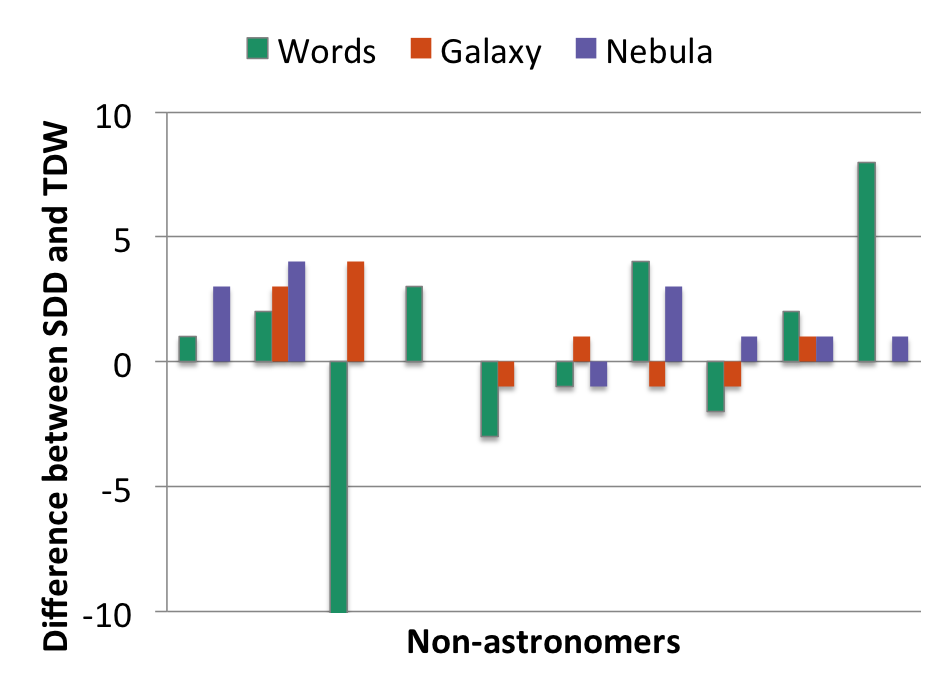}
\includegraphics[width=7.5cm,angle=0]{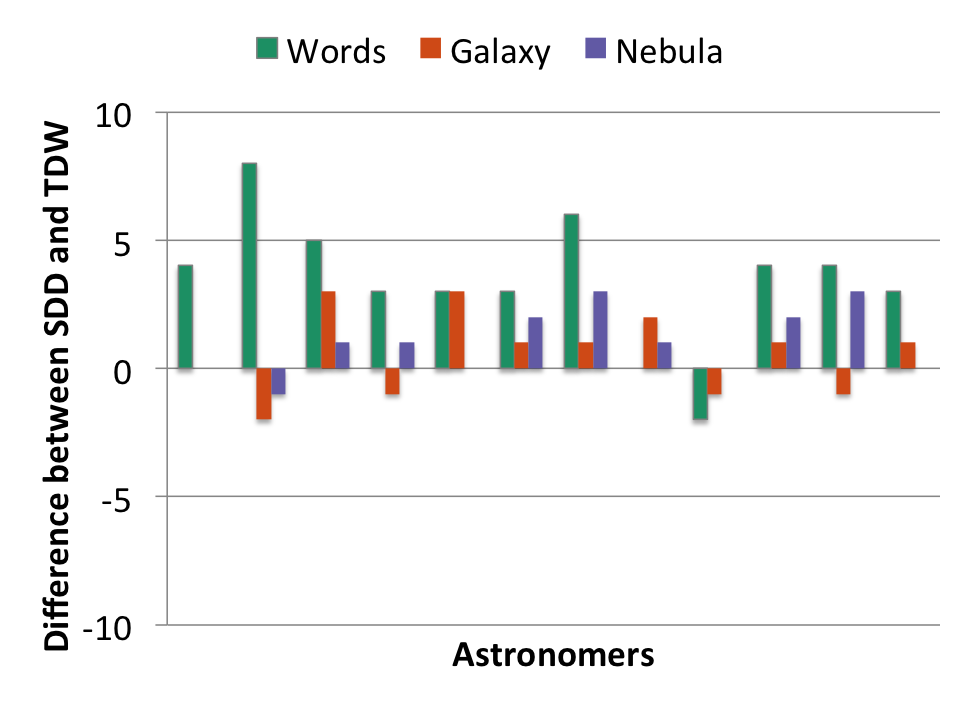}
\includegraphics[width=7.5cm, angle=0]{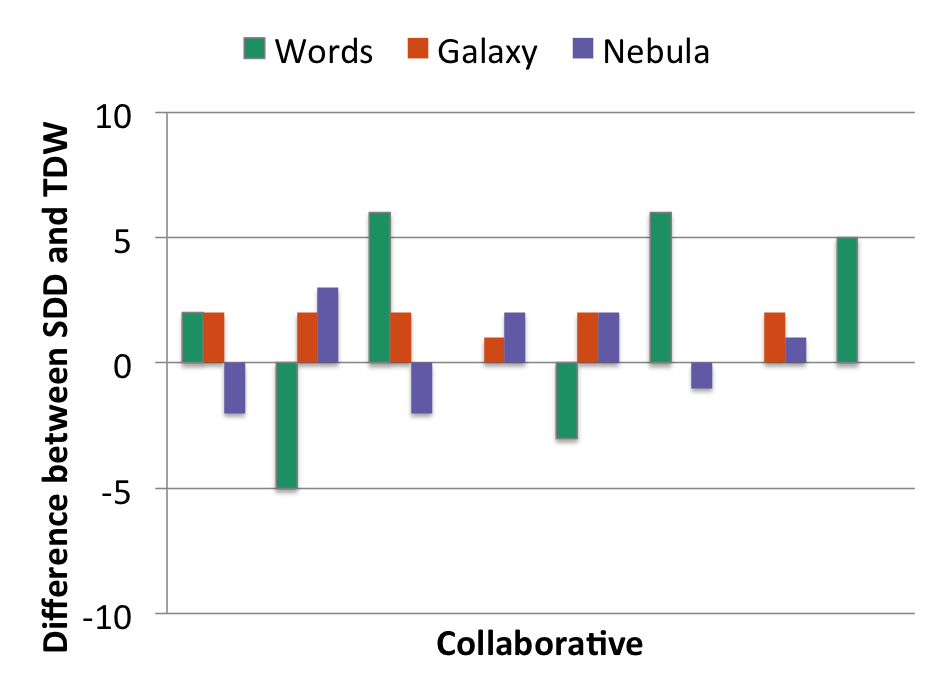}
\caption{Comparison of the success rates of the TDW and SDD for 10 non-astronomers (top panel), 12 astronomers (middle panel), and eight collaborative pairs of non-astronomers (bottom panel). Positive numbers show success favoured the TDW, negative results favoured the SDD.  The three types of experiments are the word search (green), galaxy search (red) and nebula search (purple). }
\label{fig:nonastdiff}
\end{center}
\end{figure}



\subsection{Procedure}
\label{sct:procedure}
Figure \ref{fig:exptsetup} shows the experimental set-up.   The individual target objects were presented on a 40 inch LCD TV immediately adjacent to the TDW, as well as on a laptop sitting adjacent to the SDD.  A standard Microsoft PowerPoint presentation was used to display the targets to the participant.  

As the participant identified the target (or elected to pass on finding a given target), the presentation was advanced to the next target.   Participants were given a total of 2 minutes to find as many of the targets as they could.  Once the experiment had been completed on the SDD, the participant was then shown a new set of images and targets on the TDW, again given 2 minutes for each set.  Several of the experiments were also filmed for later investigation as to how the displays were used.  

Participants were introduced to the experiments, with a brief explanation of their purpose and a demonstration of how to use the two types of displays.
The SDD was a familiar environment for all participants and very little introduction to the environment was required.  In the case of the SDD, participants were advised that they would be able to find most of the large targets without using the mouse to pan and zoom, but would need to use virtual navigation for the very small targets. The mouse operation was already second nature, though most participants attempted to minimize the mouse use, preferring to lean closer to the screen.    

Very few of the participants had ever seen a TDW before and so the experience was entirely new to them.  Those that had encountered such a display before showed little if any advantage when engaged in the structured search experiment.  The only significant advantage pre-exposure was that ability to ``zoom'' by physically approaching the TDW was already known.

In the initial experimental refinement phase, a test group of five non-astronomers was given  the same introduction to the TDW as they were to the SDD.  The result was that these participants all felt obliged to use the TDW in exactly the same way they had used the SDD, i.e. they sat well back from the screen to obtain the same field of view and used a mouse to zoom and pan rather than walk closer to the screen.  

Due to the nature of the TDW software, zooming and panning resulted in some slight image tearing as the screen refresh was not always perfectly synchronized. Moreover, the zoom was not visually active with the image jumping between zoom levels rather than scaling dynamically, as participants were used to on their SDD.  

While these issues can be mitigated with higher networking speeds, a simple alternative was found:  the participants were told that standing and approaching the screen would more effectively function as zoom (i.e. physical navigation).  This very simple training was included in the familiarization stage for the later participants.   The use of physical navigation greatly improved the user satisfaction and performance with the TDW, and presented a more realistic assessment as to how the displays should be used in practice.

For the non-astronomer singles group, all participants began the target search using the SDD.  For the astronomer group and the collaborative groups, we tested several participants with the TDW display first to see if there was any advantage to the order of exposure.  

\begin{figure}[t]
\begin{center} 
\includegraphics[scale=0.5, angle=0]{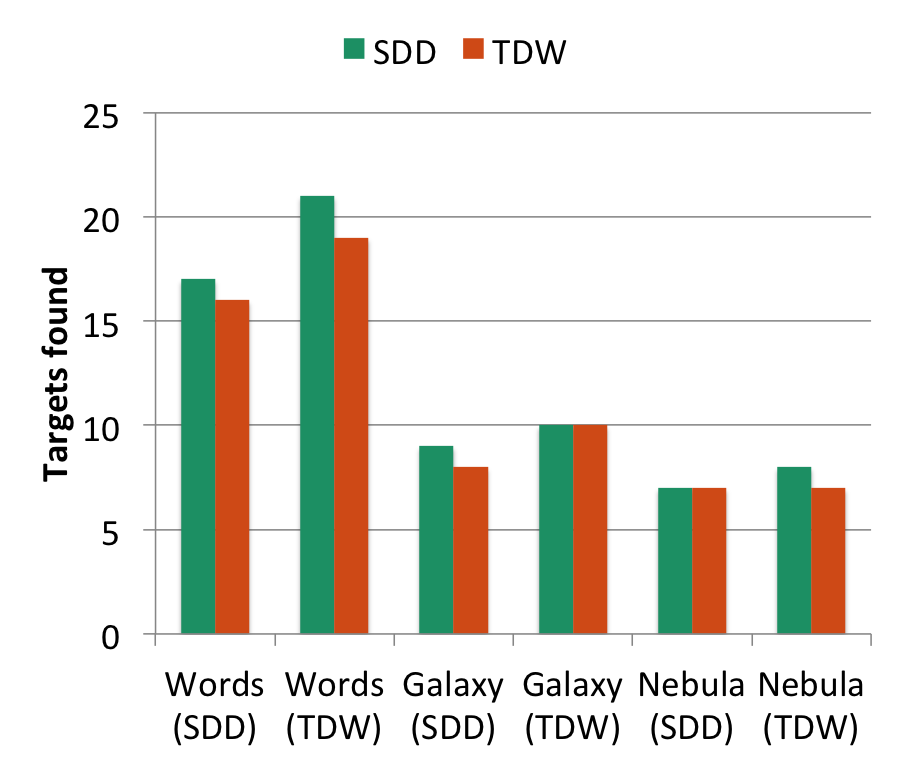}
\caption{Results based on the presentation order of the display environments.  Each pair of labelled bars indicates the image type and the first environment participants were exposed to (SDD or TDW in brackets).  The green and red bars indicate the number of targets found using the SDD and TDW respectively.  There is no strong dependence on which display technology that participants used first. }
\label{fig:presorder}
\end{center}
\end{figure}

As Figure \ref{fig:presorder} shows, there is no significant difference in performance on the target searches regardless of the order of the environments used.  The galaxy and nebula targets were alternated for the two environments in order to ensure that no advantage could be ascribed to a particular image/environment combination.  After completing the experiment in both environments, the participants were then asked to complete a survey about their experience.

\subsection{Collaborative pairs}
\label{sct:collaboration}
16 non-astronomers were asked to complete the experiment in pairs.  They received the same introduction as all the other participants, but no specific instruction was given to guide how they should share the task.  They were required to determine the best way to operate between themselves as part of the task, and in all cases settled the matter of who would operate the interface (in the case of the SDD) or how they would split the search area (in the case of the TDW), with a very brief discussion.  This process was occasionally completed during the introduction and so took no time during the task, however, in all cases it did not delay the image inspection process as the pairs began searching while discussing.  No disadvantage was observed and therefore no adjustment has been made to the performance results.  The only practical difference to the conditions of the experiment was that agreement was required from both participants for any ambiguous situation, for example, some targets could be mistaken for a similar looking non-target.  This situation included when no target could be found, at which time both participants could agree to ``pass''.

\section{Results}
\label{sct:results}
The results of the experiments are presented in the next two sections: section \ref{sct:success} shows the empirical results for the different image identification tasks and \ref{sct:survey} contains our analysis of the post-experiment survey.  We report also on the outcomes of the video observations (section \ref{sct:video}), which provide some valuable clues as to how to make better use of TDWs.  In addition, we look at the specific feedback made with regards to the TDW bezels (section \ref{sct:feedback}).

\begin{table*}[t]
\caption{Search success rates for non-astronomers (10 participants), astronomers (12 participants), non-astronomer collaborations (8 x 2 participants) and a combination of all three (10 + 12 + 8 = 30 sets of results). These results are based on the median values for the word, galaxy and nebula feature searches, with a quoted range of one standard deviation. } 
\label{tbl:success}
\begin{center}
\begin{tabular}{lrcrr}
& {\bf Group}	&$\mathbf{N_{\rm P}}$		&	{\bf SDD}		&	{\bf TDW}		\\ \hline
\multirow{5}{*}{{\bf Word search}}&Non-Astronomer 		&	$	10	$	&	$66\% \pm 2.0	$	&	$	66\% \pm 4.4	$	\\
&Astronomer 			&	$	12	$	&	$66\% \pm 1.8	$	&	$	80\% \pm 2.6	$	\\
&Non-Expert (self-rated)	&	$	15	$	&	$68\% \pm 1.7	$	&	$	72\% \pm 4.3	$	\\
&Expert (self-rated)		&	$	7	$	&	$64\% \pm 2.3	$	&	$	76\% \pm 3.0	$	\\
&Collaboration 			&		8 groups		&		$66\% \pm 3.0	$	&	$	76\% \pm 5.4	$	\\
\hline
\multirow{5}{*}{{\bf Galaxy search}}&	Non-Astronomer		&	$	10	$	&	$60\% \pm 1.7	$	&	$	75\% \pm 2.5	$	\\
&Astronomer			&	$	12	$	&	$90\% \pm 1.9	$	&	$	90\% \pm 1.3	$	\\
&Non-Expert (self-rated)	&	$	15	$	&	$70\% \pm 1.8	$	&	$	80\% \pm 2.4	$	\\
&Expert (self-rated)		&	$	7	$	&	$90\% \pm 2.1	$	&	$	80\% \pm 1.3	$	\\
&Collaboration			&		8 groups		&		$80\% \pm 0.9	$	&	$	100\% \pm 0.0	$	\\
\hline
\multirow{5}{*}{{\bf Nebula search}}&Non-Astronomer		&	$	10	$	&	$50\% \pm 2.0	$	&	$	60\% \pm 1.3	$	\\
&Astronomer			&	$	12	$	&	$65\% \pm 1.2	$	&	$	80\% \pm 1.6	$	\\
&Non-Expert (self-rated)	&	$	15	$	&	$60\% \pm 1.9	$	&	$	60\% \pm 1.9	$	\\
&Expert (self-rated)		&	$	7	$	&	$70\% \pm 1.3	$	&	$	80\% \pm 1.3	$	\\
&Collaboration			&		8 groups		& $70\% \pm 1.2	$	&	$	80\% \pm 1.6	$	\\ 
\hline
\multirow{5}{*}{{\bf Combined}}	&Non-Astronomer		&	$	10	$	&	$61\% \pm 3.3	$	&	$	67\% \pm 5.2	$	\\
&Astronomer			&	$	12	$	&	$71\% \pm 2.8	$	&	$	82\% \pm 3.3	$	\\
&Non-Expert (self-rated)	&	$	15	$	&	$67\% \pm 3.1	$	&	$	71\% \pm 5.3	$	\\
&Expert (self-rated)		&	$	7	$	&	$71\% \pm 3.4	$	&	$	78\% \pm 3.5	$	\\
&Collaboration			&		8 groups		& $70\% \pm 3.4	$	&	$	82\% \pm 5.6	$	\\ \hline
\end{tabular}
\end{center}
\end{table*}


\subsection{Success rates}
\label{sct:success}
\begin{figure}
\begin{center} 
\includegraphics[width=7.5cm, angle=0]{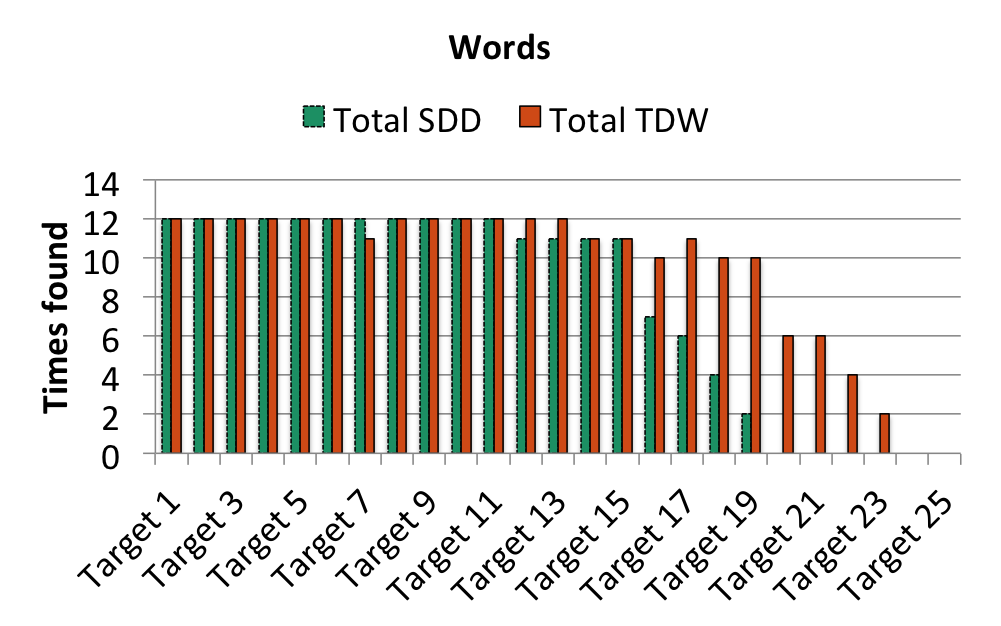}
\caption{Success rate based on individual words.  The green columns indicate the target words actually found using the SDD and the red columns are for the TDW.  Only the results from the astronomer group were used, in order to align with results shown in Figure \ref{fig:imagesuccess} as the image context was not recorded for the non-astronomer group.}
\label{fig:wordsuccess}
\end{center}
\end{figure}

\begin{figure*}
\begin{center} 
\includegraphics[width=7.5cm, angle=0]{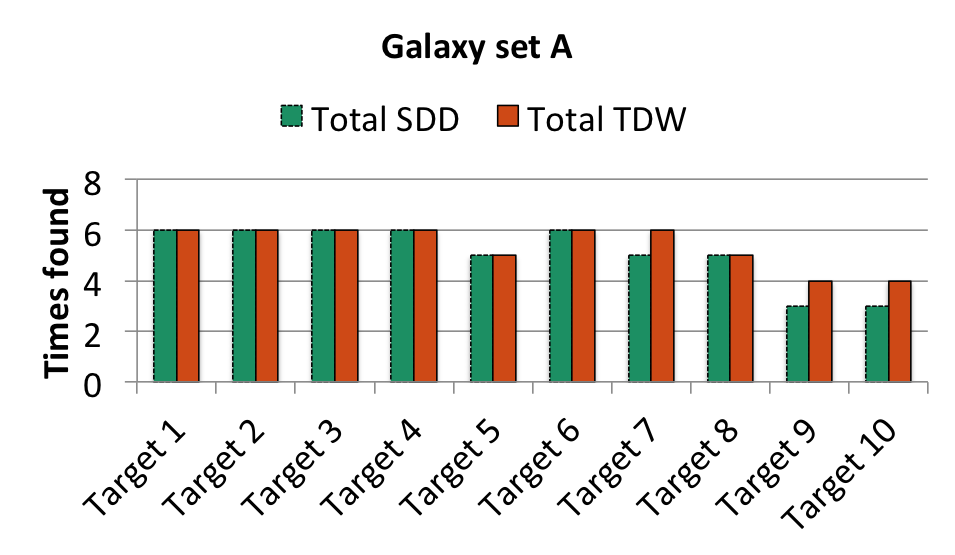}
\includegraphics[width=7.5cm, angle=0]{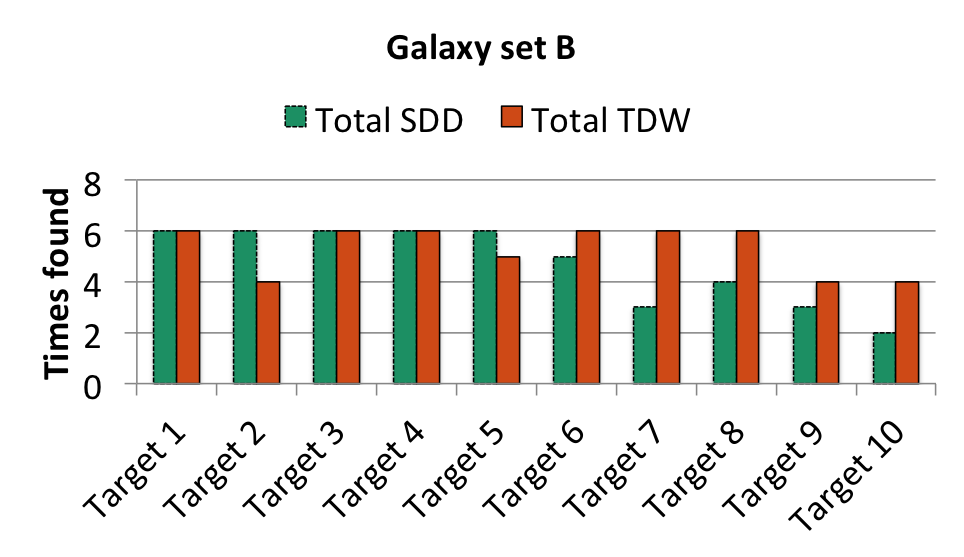}

\includegraphics[width=7.5cm, angle=0]{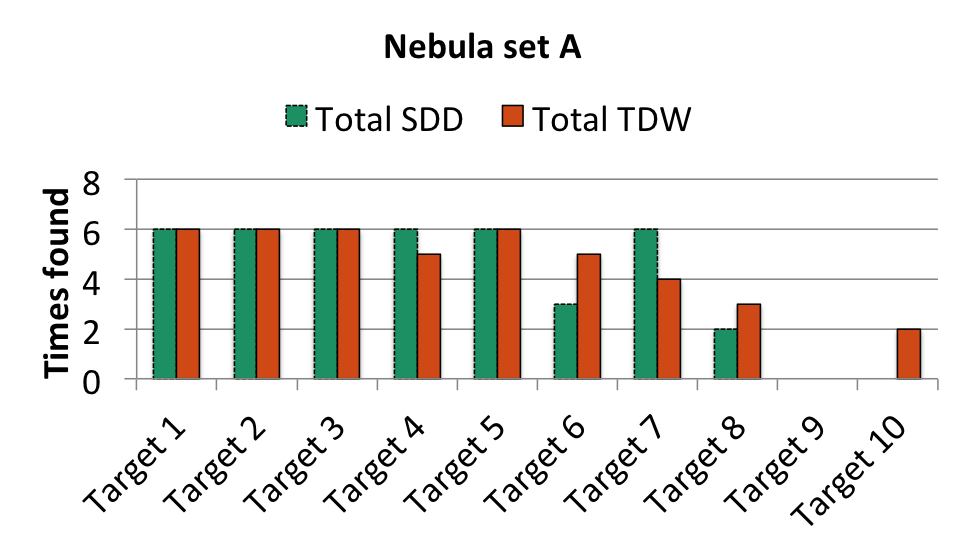}
\includegraphics[width=7.5cm, angle=0]{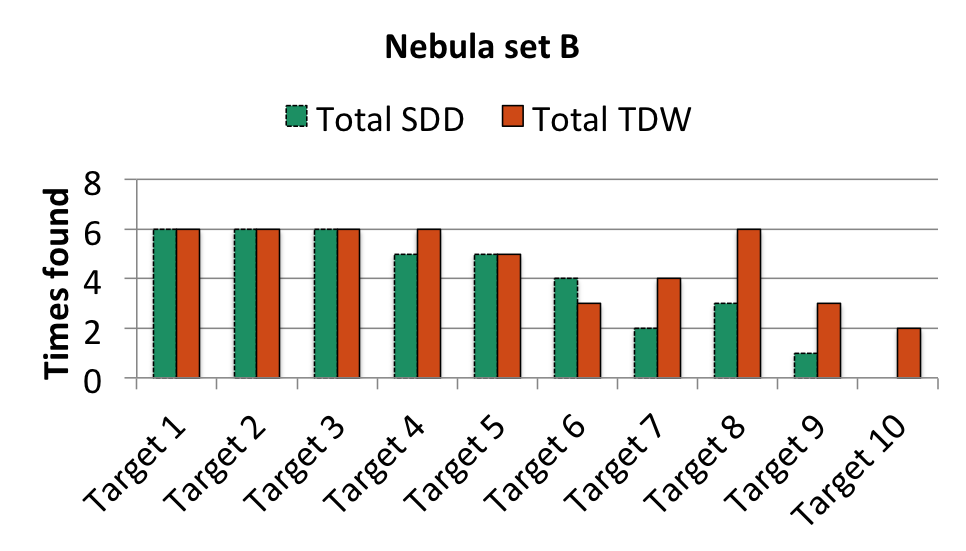}
\caption{Success rates based on individual search targets with galaxy and nebula images.  The green columns indicate the targets actually found using the SDD and the red columns are for the TDW. Only the results from the astronomer group were used as the image context was not recorded for the non-astronomer group.}
\label{fig:imagesuccess}
\end{center}
\end{figure*}

Figures \ref{fig:wordsuccess} and \ref{fig:imagesuccess} show the individual targets successfully identified. Figure \ref{fig:wordsuccess} shows the first 15 target words were found by most participants.  This corresponds to point sizes of 1000, 300 and 100, which were easily readable in both environments.  At point size 30, the words were no longer readable on the SDD and therefore virtual navigation was necessary, causing a performance decline.  This can be seen by the rapid drop in the SDD success.  Very few of the 10 point words were found in the SDD environment.  However, the TDW success rate shows only a slight decline for the 30 point words and a steady decline for the 10 point words.

Figure \ref{fig:imagesuccess} shows that the galaxy images in set A were fairly well matched for success in both environments, while set B showed a higher success rate for the TDW targets.  Similarly for the nebula sets, with the exception of target 9 in the Nebula Set A, which was not found in either environment.  As can be seen in Figure \ref{fig:nebtargetsA}, target 9 was a subset of target 2, but not particularly more difficult than target 9 of Figure \ref{fig:nebtargetsB}, also within target 2 for that set.

Table \ref{tbl:success} shows the combined success rate for each group in each environment, where success refers to the number of targets identified during the test.   These results indicate that generally performance on the TDW is slightly better than for the SDD for the same set of tasks, with the notable exception that self-rated experts actually performed slightly worse on the TDW for the Galaxy search.  However there are other factors to consider.  For example, the sample size is fairly small and the task is not necessarily indicative of typical astronomy work.  

Table \ref{tbl:success} also shows that the attempt to establish a consistent baseline between the cohorts was effective.  In the word search on a SDD, where little experiential value could be ascribed, all groups achieved very similar results.  Here the targets and the navigation method were familiar to all subjects.  However, the non-astronomers did not experience an improvement in performance when searching for words on the TDW.  This reason for this is uncertain, but could be because the astronomers' familiarity with exploring large images translated more easily to the TDW environment.  Video observations reveal that astronomers tended to adopt methodical search strategies and were quicker to adapt their strategies to the TDW environment than non-astronomers. See Section \ref{sct:video} for more information on video observations.

A useful way to view these results is to consider the comparison of the results for the specific environments.  Subtracting the results for the SDD from the TDW produces a simple comparison of the two environments, as seen in Figure \ref{fig:nonastdiff} for the 10 non-astronomers (top panel), 12 astronomers (middle panel) and eight collaborative pairs of non-astronomers (bottom panel).
The astronomer group test results are comparable to the non-astronomer cohort.  The astronomers did demonstrate slightly better performance overall, particularly with the word and nebula search.  However, the galaxy search results show that the astronomers tended to perform equally well on both the SDD and the TDW (c.f. Table \ref{tbl:success}).  Observations supported by the video recordings show that the astronomers tended to have a more systematic approach to searching, and were less confused by targets split by the screen bezels (edges).

The TDW has often been cited as an ideal environment for research collaboration.  The bottom panel of Figure \ref{fig:nonastdiff} shows the results obtained by pairing two non-astronomers for the same task.  Table \ref{tbl:success} shows that non-astronomer collaborators match or exceed the performance of a single astronomer and show marked improvement of TDW over SDD.

Figure \ref{fig:nonastdiff} shows an interesting anomaly with one participant finding the SDD to be far better for the word search than the TDW.  In this case, the participant overlooked a 1000pt word and began to search among the smaller words.  This highlights a potential trap with a TDW in that a large object presented on such a display may be too large to see, with participant approaching the TDW and effectively eliminating their chance of recognizing the word.  This may be in part due to the way the brain recognizes words as a whole and therefore may not apply to astronomical targets.

\subsection{Survey responses}
\label{sct:survey}
The participants were asked to complete a short survey after the experiment, designed to gauge their experience using the two display environments.  Participants were asked to rate their own experience with astronomical imagery, as can be seen in Figure \ref{fig:selfrate}.

\begin{figure*}[t]
\begin{center} 
\includegraphics[width=7.5cm, angle=0]{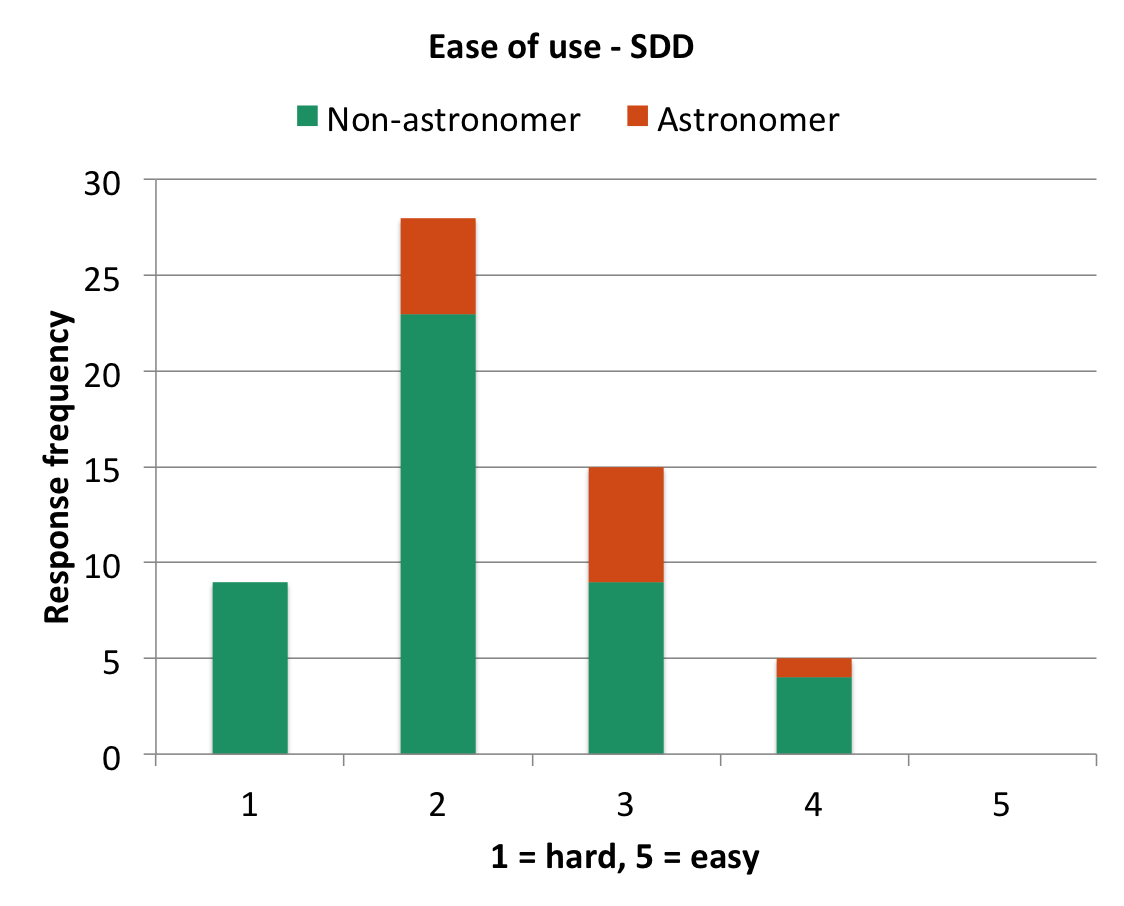}
\includegraphics[width=7.5cm, angle=0]{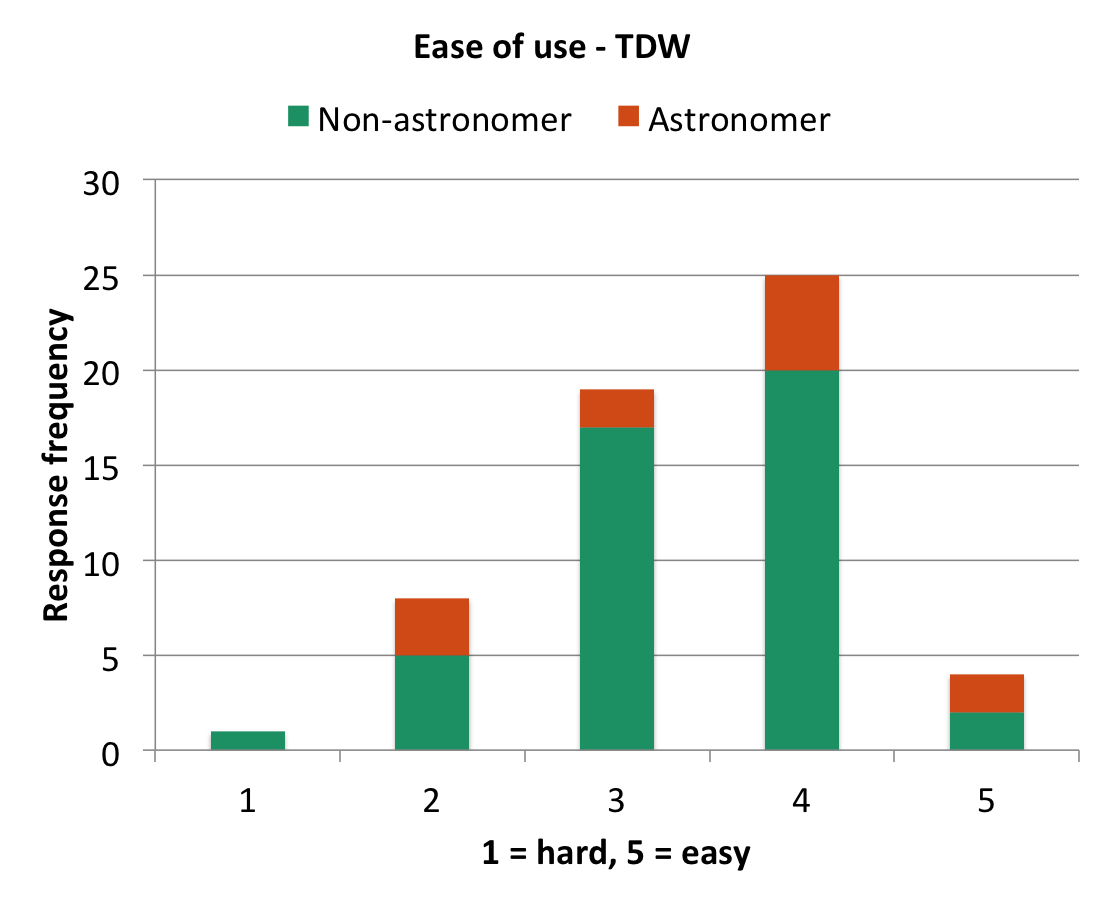}
\caption{Ease of use of the display environments. Results were obtained from post-experiment surveys completed by all 57 participants. }
\label{fig:easeofuse}
\end{center}
\end{figure*}

Figure \ref{fig:easeofuse} show the survey results for the Ease of Use of the SDD and the TDW respectively.  Results for both astronomers and non-astronomers indicate that the SDD is generally perceived as difficult for this kind of search while the TDW is generally perceived as easy to use for the same.

\begin{figure*}[t]
\begin{center} 
\includegraphics[width=7.5cm, angle=0]{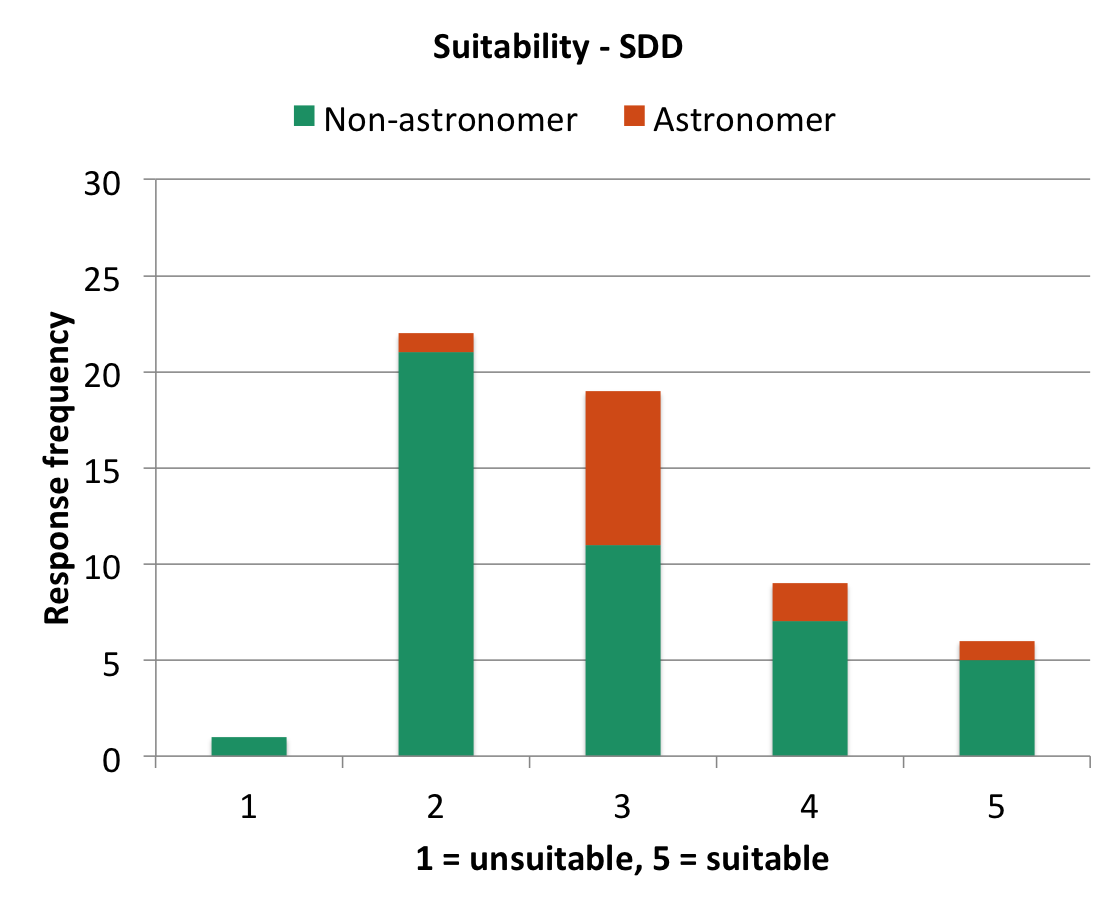}
\includegraphics[width=7.5cm, angle=0]{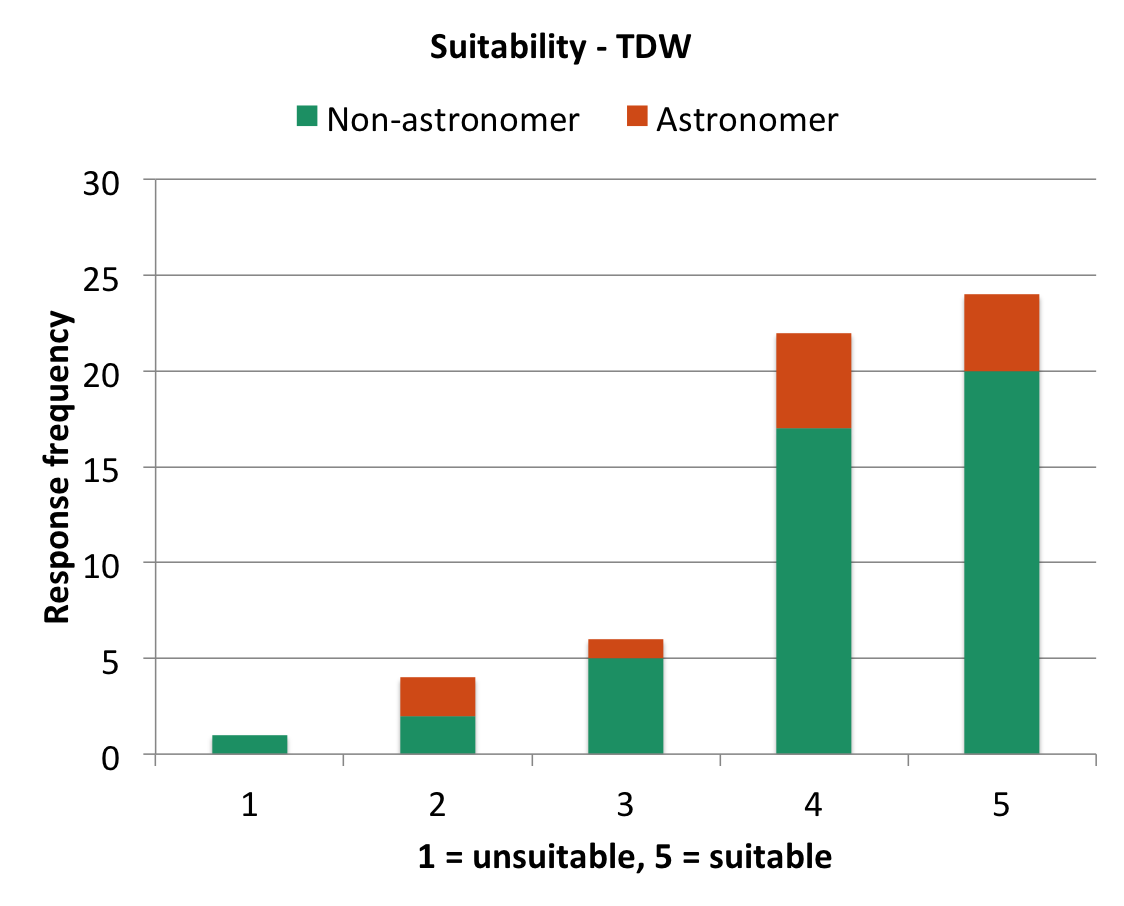}
\caption{Suitability of the display environments for target searching in ultra-high resolution images. Results were obtained from post-experiment surveys completed by all 57 participants.}
\label{fig:suitability}
\end{center}
\end{figure*}

Participants were also asked to rate the suitability of the two environments to the tasks presented.  Figure \ref{fig:suitability} show that the results for the SDD are skewed toward Unsuitable for the search tasks while the participants found the TDW was generally well suited.

\subsection{Video observations}
\label{sct:video}
Several of the participants were also filmed to record the manner in which they used the display environments.  

When using the TDW, several participants found themselves overlooking extremely simply targets, particularly in the word search, by assuming that the target they were seeking must be smaller than it actually was.  However, generally the approach to searching was fairly uniform, with the subjects standing back to get an overview of the image, and then approaching promising regions of the image.  In the case of words, the advantage to the TDW over the SDD was that even the very smallest of words could be clearly read when close, while the same target on the SDD was not even visible when zoomed to full extents.

However, the experiment was designed to make the first targets easy to find and subsequent targets were made progressively harder.  This gave a distinct advantage to the SDD for the early targets, and therefore considerably more time was available for finding smaller targets.  On the TDW, however, the larger targets were sometimes overlooked, occasionally due to the splitting of the target by the screen bezels, or due to the participants' assumption that the target must be smaller than it actually was.

In general however, the video review shows participants were more methodical in searching for small targets on the TDW than on the SDD.  It appears that the participants were more easily able to identify individual screens that they had already searched; compared to trying to remember which region they had searched of the image on the SDD.  The screens of the TDW made for a simple segmentation that was easy to remember.

Furthermore, participants working with a partner found the TDW to be naturally separated into halves with each partner being responsible for their own half.  Working together on the SDD, these participants found ceding control to someone else to be frustrating.  Some chose to share the task of controlling the mouse, alternating between tasks, while others simply directed their partner by pointing in the direction they wished to explore.  This resulted in some confusion, though some collaborations quickly settled into very effective teamwork.

On the TDW, splitting the screens into left and right did not always produce harmony.  When one participant became convinced that the target was not on their side, they began to encroach on their partner's domain.  For some, this resulted in an unspoken agreement to swap sides, while not so for others.  However, the success of the collaborations between non-astronomers produced results that were generally better than a single astronomer (c.f. Table \ref{tbl:success}).  Unfortunately there were not enough astronomers to test collaborative behaviours.

\subsection{Feedback on bezels}
\label{sct:feedback}
\begin{figure}[t]
\begin{center} 
\includegraphics[width=7.5cm, angle=0]{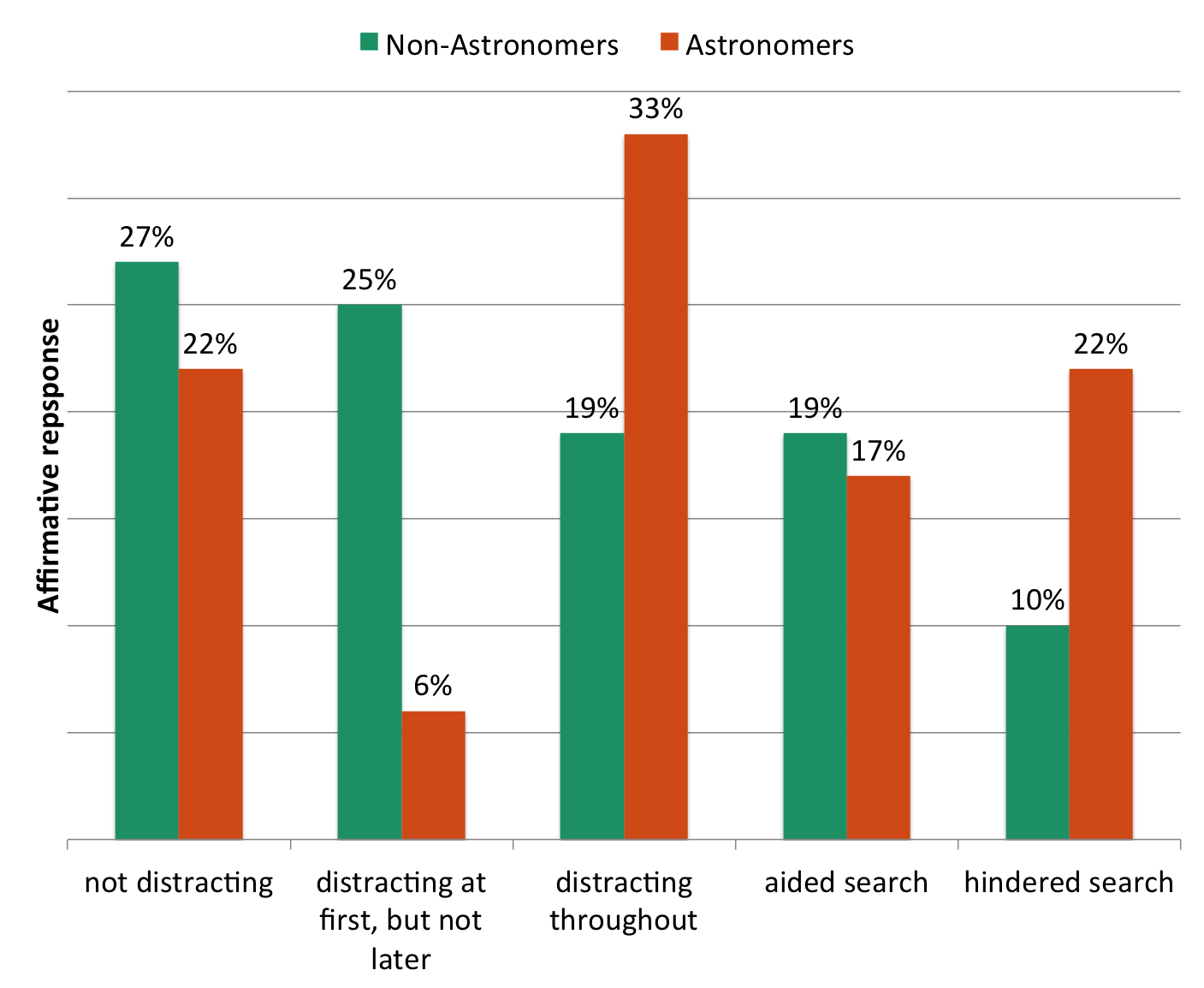}
\caption{Impact of screen elements on the task based on post-experiment survey.  The bars indicate the percentage of non-astronomers (green) and astronomers (red) who provided an affirmative response to questions regarding the level of distraction caused by screen elements, and their perceived impact on the search process. }
\label{fig:interference}
\end{center}
\end{figure}

Participants were also asked to comment on the structure of the TDW and how the ``screen elements'' impacted on the subjects' ability to complete the task.  We defined screen elements as anything that interrupted the subjects' view of the search image.  This included the SAGE icons and toolbars and the screen bezels.  The results can be seen in Figure \ref{fig:interference}.

For the non-astronomer group, $19\%$ found the screen elements were a continuous distraction while $25\%$ found them initially distracting but quickly learned to ignore them.  $27\%$ found the screen elements did not distract them at all during the tasks.  In fact, $19\%$ of the non-astronomer respondents felt the bezels actually aided their search, compared to $10\%$ who felt they made the task more difficult.

The astronomer group found the screen elements to negatively impact on their experience more so than the non-astronomer group.  $33\%$ found the screen elements distracting throughout the experiment, while $6\%$ found them initially distracting, but not so later.  However, much like the non-astronomer group, $22\%$  found the screen elements did not distract them from searching for their targets and $17\%$  found the bezels aided their search strategy. However, $22\%$ found the screen elements hindered their efforts.

The general comments relating to the display environments and the screen bezels generally support the results described above.  For the SDD, the comments suggest that the environment was well suited to examining very large images in a broad context, when the entire image could be seen.  However, small details within these large images were much harder to find and the context was often lost, making strategic searches harder.  

Contrasting this with the TDW, the expanse of the display itself sometimes made observing the whole image harder as the participant needed to be much further back to achieve the same field of view, or turning the head considerably more.  However, the dynamic nature of physically approaching the screen and being able to see extremely fine detail within the large image made the search for small targets easier on the TDW.  This combined with the physical break in the image due to the screen bezels provided subjects with an easier search methodology.

\section{Discussion}
\label{sct:discussion}
Previous experiments have shown that for searches of small targets within much larger images, a distinct advantage exists for a TDW \citep{ball2005b, ball2007, yost2007}, when the case for  physical navigation being preferable to virtual navigation is clear. However, when the target size varies, the advantage is less apparent, though the case for physical versus virtual navigation remains.  This is because the larger targets can be found with relative ease in a large image, even when it is presented subsampled on a SDD.  

Our study showed that participants typically attempted to gain an overview of the whole image to identify regions of interest.  In the case of a large target, this was often more readily found on the SDD as it was quicker to obtain this overview and therefore ascertain the target.  As we chose targets that would not easily be confused with other objects, this continued to be the case as the targets got smaller as the participants were able to recognize the approximate shape and zoom quickly only on that part of the image.  However, as the targets became too small to even approximately identify, virtual navigation became essential and performance (i.e. success rate) declined rapidly.

While the results show a slight advantage to the TDW for the target searches, it was not as significant as expected.  This is in part because the experiment deliberately spanned a range of difficulty, and thereby is inclusive of both the SDD and TDW advantages.  However, results obtained from the post experiment survey indicate that participants decidedly preferred the TDW experience over the SDD, even if their performance results did not show a marked difference.  Indeed, several participants believed themselves to have performed better with the TDW when they had in fact performed better with the SDD.  

The novelty of the TDW environment cannot be ignored and the fact that observing very large images in such an immersive environment will have had some emotional impact on the way participants viewed the experiment.  Also, the experiment was clearly investigating the perceived value of TDWs compared to the SDD, and may have skewed participants' perception in favour of the more novel technology.  A cross-section of participants primarily sourced from universities may reflect that preference for new technology.  However, the primary use of the TDW is in this sector and therefore the performance of such a cohort remains relevant.

The very fact that participants preferred the TDW environment is important even when it did not correspond to increased performance.  This suggests that participants might be more inclined to persist with the TDW environment further than with the SDD, however this may be a result of the novelty factor.

No matter how much the novelty factor plays a part, we find that the experience of the participants in this study reflect the results of previous experiments that show performance improvement when virtual navigation can be avoided.

The results from this study indicate considerable opportunities for further work in general testing of TDWs and domain specific testing.  This experiment used specifically constructed conditions to examine aspects of the display environments, however, these conditions aren't necessarily indicative of typical astronomical activities.  Therefore, before astronomers are likely to include a TDW into their workflow, there needs to be evidence that TDW will actually improve efficiency and/or reduce errors or omissions.

Based on our understanding of how participants used the displays, and in response to individual comments, we identify several areas for future investigation:
\begin{itemize}
\item {\bf Multiple image inspection.} 
Our experiments focused on one particular use case for TDWs - inspection of individual ultra-high resolution images.  However, there is an alternative way to take advantage of the display pixels: instantaneous display of many individual lower-resolution images.
Consider the simple case of classifying structures in images: a crucial skill in many fields of astronomy which remains difficult to fully automate.  Using a TDW would allow astronomers to maintain a view of many classified structures, thereby assisting the evaluation of unclassified structures.  This would provide an opportunity for refinement and reclassification, as previous decisions are still available for scrutiny.  
\item {\bf Extended exposure.} The experiments in this study were designed to run within a 30-minute period to make it easier for participants to commit their time.  The results from the survey indicate that most participants felt the TDW was easier to use than the SDD, as well as being more suitable to the task.  This suggests that extending the exposure time in a variety of ways might yield more distinguishing results.  For example, if the experiment were not time limited, how long would it take to find all targets?  Alternatively, if the participants were to use the TDW each day for a period of time, for example, a week, would they improve their performance compare to a control group using SDDs?  
\item {\bf Features retained over extended periods.}
Following on from the previous item, it would be useful to learn whether a TDW aids in the recall of multi-scale features in ultra-high resolution images.  Such an experiment would test a participant's ability to relocate features in an image that they had previous been able to find, or had been shown to them.  This experiment would look at the difference between short and mid-term memory to see if the TDW exposure shows a difference compared to SDD exposure.
\item {\bf Collaborative exploration.}
Our study looked at a very basic form of collaboration with two participants working together to share the task.  However, there are several variations that would be worthy of further investigation.  For example, rather than sharing a SDD, can the frustration caused by sharing control be alleviated by providing each participant with a SDD?    It would be expected that more overlap of searched area would occur, but this might be mitigated if each participant could observe the other's display.

Moreover, how exactly does communication between participants occur in this situation, either naturally or guided?  Is physical proximity necessary, or is virtual proximing via teleconferencing facilities sufficient?  Such a study has added relevance for the case where participants are working off physically remote but linked TDWs, as in the case of OptIPortals.  Finally, increasing the number of participants beyond two, might establish a relationship between screen size and practical use with respect to the number of people observing that data. 
\item {\bf Consumer 4K UHD displays.} With the recent availability of consumer-grade 4K UHD displays, it would be valuable to repeat the experiment in an environment that might represent an effective combination of the SDD and TDW.  While not providing the number of pixels available on a TDW, the advantages of a SDD would be brought to bear, and might produce a cost-effective compromise.  Depending on the screen size, this might also prove to be a viable collaborative environment as well as being suitable for an individual.  While software like SAGE would work with a 4K display, a significant benefit of running a display from a single computer is that windowing environments can be configured easily and no inter-machine synchronisation is required.  This means that all applications can be run without modification. 
\end{itemize}

\section{Conclusion}
\label{sct:conclusion}
The amount of information captured by current and future astronomical instruments greatly outstrips the resolution of both current and on-the-horizon displays.  TDWs provide a cost-effective method of achieving an order of magnitude increase in display resolution, thereby enhancing the presentation of astronomical data and potentially optimising the consumption of information. However, the notion that TDWs are essential when dealing with extremely large images is not so clear.


The results from this study indicate that TDWs provide a better platform for searching for discrete targets within large images than with a SDD.  It also shows that astronomers perform somewhat better than non-astronomers at extracting information from extremely large images (likely due to their more systematic approach to searching), and that the collaborative combination of two non-astronomers using a TDW rivals the experience of an individual astronomer. However, the study also indicates that there is a great variety of aptitude of participants, suggesting that TDWs might greatly enhance the performance for some individuals, while providing little help to others.  It also shows that the type of content also has a significant impact on the participants ability to identify targets.

This experiment has borne out the results of earlier research highlighting the benefits of physical versus virtual navigation, and the value of TDWs for searching for very small targets.  However, in astronomy, as in many other disciplines, there is a great variation in the physical scale and the ``visual connectivity'' of the objects to be studied.  Our experiments highlight the differences between looking at words, which are processed differently by the brain, compared to identifying isolated galaxies in wide field surveys or visually continuous nebula.

A TDW provides a very impressive environment to examine images and participants enjoyed the experience, which significantly influences their perception of the suitability of the TDW environment.  While such value is difficult to quantify, it suggests that the availability of a TDW can be a useful addition to the astronomer's work flow - if only because using one is a more enjoyable task than being seated at the desktop.  We are encouraged to believe that the TDW has now come-of-age for astronomy, particularly as a collaborative environment.  Ultimately, the practicality of the wider up-take of TDWs for astronomy is contingent on increasing ease-of-use (e.g. through the SAGE environment), suitable interface options and simple training in the use of physical navigation.


\section*{Acknowledgments} 
We thank the participants for contributing their time to this project. All experimental work was approved and conducted in  accordance with the requirements of Swinburne University’s Human Research Ethics Committee (SUHREC). We also thank Dr Simon Cropper and Dr David O'Conner (Psychology dept., University of Melbourne) for their advice on the experiment design, and Christoph Willing (University of Queensland) and Luc Renambot (University of Chicago) for their technical advice in the use of the SAGE environment. We also thank ITS Research Services at the University of Melbourne for the use of the OzIPortal for this experiment.

\appendix
\section{The OzIPortal Project}
\label{sct:oziportal}

We now look at the OptIPortal project and a specific TDW, the OzIPortal, in more detail in order to understand some of the reasons why these devices have not already become standard elements of the research workflow across diverse scientific disciplines.

\subsection{The OptIPortal Project}
\label{sct:optiportalproject}
The OptIPortal project grew out of the Optiputer project, a US government funded project to connect high performance computing facilities together via optical networks, called Lambdas \citep{defanti2009, taesombut2006}.  With such powerful data processing and transfer resources in the background, an enhanced visualization capability was required.   This was initially achieved using several projectors, with edge-blending techniques to compensate for luminosity fall-off between adjacent projections.  The OptIPortal project took this methodology to the next level, adopting high-resolution off-the-shelf displays to create tiled surfaces.  Software was developed to make the management of the TDW relatively transparent so that the users could focus on the research content \citep{defanti2009}.  

Collaborative environments such as the OptIPortal network, were designed to allow distributed research teams to work together simultaneously on the resulting imagery and analysis.  This would provide greater opportunity for collaborative research, leading to greater understanding of the data \citep{smarr2009, sims2010, yamaoka2011}.

While the project promised a simple, powerful and interconnected system, there were many problems with the early incarnations of OptIPortals, primarily due to immaturity of the associated software.  To tackle this, workarounds are commonly employed, such as manual data processing steps.  Whilst giving the appearance of success for visualization, this led to some misunderstandings as to what an OptIPortal actually is and its overall utility as part of research workflows.


\subsection{The OzIPortal Experience}
\label{sct:oziportalexp}
In 2008, The University of Melbourne launched the OzIPortal, a 98 megapixel TDW, with considerable fanfare.  It was lauded as an amazing research tool:
"In an Australian first, this next-generation platform set to revolutionize the way Australia interacts with the rest of the world allows real-time, interactive collaboration across the globe, combining high-definition video and audio with the sharing of ultra-resolution visualizations from a broad range of disciplines." \citep{calit2web}.

Despite the high level of interest generated by early demonstrations, the OzIPortal failed to attract a significant commitment from the research community.  

The OzIPortal was configured using 24 x 2560x1600 LCD displays, initially in an 8x3 arrangement and later in 6x4.  12 slave nodes with dual-head graphics cards were used to drive two monitors each.  Reducing the number of slave nodes improved the operation of the TDW without reducing the performance.  Data was exported from the head node via NFS to each of the slave nodes.  An additional machine was required to provide a real-time video stream that would allow any video signal captured via HDMI to be presented on the TDW.  In this way, the OzIPortal was able to include low latency, high-definition video conferencing content alongside other stored content.

One of the initial drivers for the OzIPortal was to establish a dedicated gigabit link from The University of Melbourne through to CALIT2 in the United States \citep{oziportalnewsweb}.  This was successfully implemented on a layer 2 network via AARNet. The purpose of the network was to demonstrate the rapid transfer of massive datasets, to show how distant collaborators could work on the same datasets at the same time.  In practice however, it was necessary to distribute the data in advance, as some of the data was too large to deliver in a timely manner, even over the dedicated link.  Instead, the link was used primarily to stream uncompressed video directly to the TDW, instead of via a video conference codec.
Finally, it took several technicians to satisfactorily operate the TDW, and a great deal of testing beforehand was required to minimize disruptions during events.  Day-to-day operations could not be sustained with such human resource demands and as such provided a less than satisfactory experience for users.

Ultimately, what was required for the successful deployment of a TDW in the research workflow was a more stable system, that was easier to use, did not require high-level of support, automated preparation of content, and the ability to run applications specific to individual scientific disciplines.   Our recent experiences with the OzIPortal, particularly through the use of the SAGE environment, is a positive step towards these outcomes.

\end{document}